\documentclass{elsarticle}
\usepackage[utf8]{inputenc}
\usepackage{graphicx}
\usepackage{placeins}
\usepackage{amsmath}
\usepackage{bm}
\usepackage{tikz}
\usepackage[ruled,vlined]{algorithm2e}
\graphicspath{{pictures/}}
\usepgflibrary{arrows}
\usepackage{rotating}  
\usepackage{comment,xspace}
\usepackage{hyperref}

\newproof{pf}{Proof}

\begin{document}

\begin{frontmatter}
\title{Modified Bee Colony optimization algorithm for computational parameter identification for pore scale transport in periodic porous media}

\author[nefu]{Vasiliy V. Grigoriev}
\ead{d\_alighieri@rambler.ru}

\author[itwm,ban]{Oleg~Iliev\corref{cor}}
\ead{iliev@itwm.fraunhofer.de}

\author[nsi,rudn]{Petr N. Vabishchevich}
\ead{vabishchevich@gmail.com}

\address[nefu]{Multiscale Model Reduction Laboratory, North-Eastern Federal University, 58 Belinskogo st, Yakutsk, 677000, Russia}
\address[itwm]{Fraunhofer Institute for Industrial Mathematics ITWM,  Fraunhofer-Platz 1, D-67663 Kaiserslautern, Germany}
\address[ban]{Institute of Mathematics and Informatics, Bulgarian Academy of Science, Sofia, Bulgaria}
\address[nsi]{Nuclear Safety Institute, Russian Academy of Sciences, 52, B. Tulskaya, Moscow, Russia}
\address[rudn]{RUDN University, 6 Miklukho-Maklaya st, Moscow, 117198, Russia}

\cortext[cor]{Corresponding author}

\begin{abstract}

This paper discusses an optimization method called Modified Bee Colony algorithm (MBC) based on a particular intelligent behavior of honeybee swarms. The algorithm was checked in a few benchmarks like Shekel, Rozenbroke, Himmelblau and Rastrigin functions, then was applied to parameter identification for reactive flow problems in periodic porous media. The simulation results show that the performance and efficiency of MBC algorithm are comparable to the other parameter identification methods and strategies, at the same time it is able to better capture local minima for the considered class of problems. The proposed identification approach is applicable for different geometries (random and periodic) and for a range of process parameters. In this paper the potential of the approach is demonstrated in identifying parameters of Langmuir isotherm for low Peclet and high Damkoler numbers reactive flow in a 2D periodic porous media with circular inclusions. Finite element approximation in space and implicit time discretization are exploited. 
\end{abstract}

 
\end{frontmatter}

\section{Introduction}
Among the tasks of continuous finite-dimensional optimization, the most important from a practical point of view and, at the same time, the most difficult is the class of problems of global optimization. Methods for solving the global optimization problem are divided into deterministic stochastic and heuristic methods. Heuristic methods are a relatively new and rapidly developing class of global optimization methods. Among these methods, evolutionary and behavioral methods stand out. Behavioral methods are multiagent methods based on modeling the intellectual behavior of agent colonies (swarm intelligence). In nature, such an intelligence is possessed by groups of social insects, for example, termite colonies, ants, bees, and some species of wasps.
The dynamics of the population of social insects is determined by the interactions of insects with each other, as well as with the environment. These interactions are carried out through various chemical or physical signals, for example, pheromones secreted by ants. 

Swarm intelligence has become a research interest to many research scientists of related fields since 2005. In general, there is quite a lot of modifications of the algorithm. First of all, the Bees algoritm starts by Pham, Ghanbarzadeh et al. in 2005 \cite{pham2005bees,pham2006bees}. It mimics the food foraging behaviour of honey bee colonies. More about effectiveness and specific abilities of this algorithm have been proven in the following papers \cite{pham2009bees,pham2014benchmarking,pham2015comparative}. Artificial Bee Colony (ABC) algorithm was introduced by Karaboga in 2005 \cite{karaboga2005artificial}. This algorithm is a swarm based meta-heuristic algorithm. More about this algorithm like performance and convergence can be found in the following papers \cite{karaboga2008performance,karaboga2009comparative}. The main difference of ABC from usual Bee Colony algorithm is that bees can leave their region if there is not enough nectar. This can be very useful when looking for a global extremum: the algorithm can converge much faster. The undoubted advantage of this algorithms is that it can be generalized to problems of any dimension, also it can be easily be run in parallel on supercomputers.

In this work we will use Modified Bee Colony algorithm (MBC) because the ABC is suitable only for finding a global extremum, in our problem, due to noise in the experimental data, local extremums can occur and it is important for us to know their coordinates. Therefore, MBC algorithm is more preferable to us as a tool. The difference from usual Bee Colony is that if we are unable to improve our found extremum for some time, we shrink the local search area.

To evaluate the operation of the algorithm written in the Python, test runs were carried out to find global and local minimums on well-known test functions. Despite the fact that all the functions below are generalized for the multidimensional case, in this work we will run our program only for the two-dimensional case, since the algorithm itself can easily be rewritten for any dimensions, and in the two-dimensional case it is easier to clearly demonstrate the operation of the algorithm itself. As benchmark problems were choosen Shekel, Rosenbrock, Himmelblau and Rastrigin functions.

The reactive transport in porous media is important component of many industrial and environmental problems like water purification, soil pollution and remediation, catalytic filters, $CO_2$ storage, oil recovery, etc., to name just a few. Historically, most of the theoretical and experimental research on transport in porous media in general, and on reactive transport in particular, has been carried out at macroscopic Darcy scale \cite{bear2013dynamics,helmig1997multiphase}. In many cases, the bottleneck in performing computational modeling of reactive transport is the absence of data for the pore scale adsorption and desorption rate (or in more general, the parameters of the heterogeneous reactions). Despite the progress in developing devices to perform experimental measurements at the pore–scale, experimental characterization of these rates is still a very challenging task. 

In the case of heterogeneous (surface) reactions at pore scale, the species transport is coupled to surface reaction via boundary conditions. When the reaction rates are not known, their identification falls into the class of boundary value inverse problems, \cite{lavrentev1986ill,alifanov2011inverse,isakovinverse,samarskii2007numerical}. The additional information which is needed to identify the parameters is often provided in form of dynamic change of the  concentration at the outlet (e.g., so called breakthrough curves).  In the literature, inverse problems for porous media flow are discussed mainly in connection with parameter identification for macroscopic, Darcy scale problems. An overview on inverse problems in groundwater Darcy scale modeling can be found in  \cite{sun2013inverse}. Identification of parameters for pore scale models is discussed in this paper, and the algorithms from \cite{sun2013inverse} and other papers discussing parameter identification at macroscale can not be applied here without modification. Let us shortly mention some general approaches for solving inverse problems.  

Different algorithms can be applied for solving parameter identification problems, see, e.g. 
\cite{tarantola2005inverse,Aster2013}. Many of the algorithms exploit deterministic methods based on Tikhonov regularization technique  \cite{tikhonov1977solutions,engl2014inverse} and target at minimizing a functional of the difference between measured and computed quantities. An important part of such algorithms is the definition of feasible set of parameters on which the functional is minimized. Local or global optimization procedures are used in the optimization \cite{horst2013handbook,nocedal2006numerical}.
In this sense it could be pointed out that there is certain similarity between the mathematical formulation of an optimization problem and of a parameter identification one.

Stochastic-deterministic methods are also popular approach for solving parameter identification problems. A variant of the method based on deterministic sampling of points looks appropriate for the topic considered here. A stochastic approach for global optimization in its simplest form consists only of a random search and it is called Pure Random Search \cite{Zhigljavsky2008}. In this case the residual functional is evaluated at randomly chosen points from the feasible set. 
Sobol sequences \cite{sobol1976uniformly,sobol1979systematic} can be used for sampling. The sensitivity analysis tool SALib \cite{Herman2017} has shown to be appropriate tool for this. Such an approach is successfully used, for example, in multicriteria parameter identification \cite{sobol1981choosing}. 

The solution of the inverse problems we are interested in, is composed of two ingredients: (multiple) solution of the direct (called also forward) problem, and the parameter identification algorithm.

The goal of this paper is to contribute to the understanding of the formulation and the solution of a class of parameter identification problems for pore scale reactive transport in the case when the measured concentration of the specie at the outlet of the domain is provided as extra information in order to carry out the identification procedure. Deterministic and stochastic parameter identification approaches are considered. The influence of the noise in the measurements on the accuracy of the identified parameters is discussed. Multistage identification procedure is suggested for the considered class of problems. The proposed identification approach is applicable for different geometries (random and periodic) and for a range of process parameters. In this paper the potential of the approach is demonstrated in identifying parameters of Langmuir isotherm for low Peclet and low Damkoler numbers reactive flow in a 2D periodic porous media with circular inclusions. It is supposed that this paper is the first one in series of papers dedicated to this topic. Simulation results for random porous media and other regime parameters are subject of follow up papers. 

The reminder of the paper is organized as follows. The Modified Bee Colony Algorithm is described in detail in Section 2. The direct problem is considered in Section 3. At pore scale, single phase laminar flow described by incompressible Stokes equations, and solute transport described by convection-diffusion equation, are considered. The surface reaction is accounted in the boundary conditions. Henry and Langmuir adsorption isotherms are considered here \cite{kralchevsky2008chemical}, identification is carried out for  adsorption and desorption parameters in the Langmuir isotherm. Section 4 is dedicated to description of the used computational algorithm. Finite element method is exploited after triangulation of the computational domain. The numerical investigation of the grid convergence and sensitivity with respect to parameters is also presented in this Section. The set up of the parameter identification problem for reactive flow in porous media is described in Section 5. Finally, Section 5 summarizes the results presented in this paper.

\section{Bee Colony Algorithms}

\subsection{Modified Bee Colony Algorithm}

In this work we will use Modified Bee Colony algorithm (MBC) because the ABC is suitable only for finding a global extremum, in our problem, due to noise in the experimental data, local extremums can occur and it is important for us to know their coordinates. Therefore, MBC algorithm is more preferable to us as a tool. The difference from usual Bee Colony is that if we are unable to improve our found extremum for some time, we shrink the local search area. Let us describe the idea of this algorithm. There are scout bees $sb$ that do random searches. After exploring the area, judging by the amount of nectar found, the area is divided into several local areas. The first $n$ areas are called the best locations to which more agent bees $abb$ are sent, to the remaining locations less agent bees $abp$ are sent. All bees working in their locations can search around their area to increase the amount of nectar. If the bees cannot improve the result for a certain time $\tau$, then they narrow their search.

In general, our implementation of the algorithm requires 8 control parameters, including
\begin{itemize}
\item[-] best locations $n$\;
\item[-] perspective locations $m$\;
\item[-] half side of local area $d$\;
\item[-] Euclidean distance value $\delta$\;
\item[-] number of scout bees $sb$\;
\item[-] number of agent bees at the best locations $abb$\;
\item[-] number of agent bees on perspective locations $abp$\;
\item[-] parameter of the number of iterations during which it will not be possible to improve the result, after which the zone will decrease $\tau$\;
\end{itemize}

The algorithm is listed below:\\
\begin{algorithm}[H]
\SetAlgoLined
 Initialization of scout bees\;
 Sort them by nectar value\;
 Calculating the Euclid distance and divide into regions\; 
 \For{Check all regions}{
 \If{Iteration counter >= Maximum iteration}
 {Break\;}
 Set initial extremum\;
 Set \textit{divider} to zero\;
 Set local search area size\;
 Initialize local area where extremum is in the center\;
 \While{Until converge}{
  \textit{Iteration counter += 1}\;
  Initialize agent bees\;
  Sort them by feed value\;
  \eIf{New found extremum is better than previous one}{
   Change extremum to new one\;
   Move local search area to the new extremum\;
   Set \textit{fail = 0} \;
   }{
   \textit{fail += 1}\;
  }
  \If{\textit{fail} == \textit{stop fail}}{
   Set \textit{fail = 0}\;
   Set \textit{divider} += 2\;
   Shrink local search area \textit{divider} times\;
    \If{Euclid distance through founded extremums <= $\varepsilon$}{
    Break\;}
   }{
   }
  \If{Iteration counter == Maximum iteration}
  {Break\;}
  }
 }
 \For{All regions}
 {
 Print found extremum coordinates and values
 }
\caption{Modified Bee Colony algorithm}
\label{algorithm}
\end{algorithm}

\subsection{Benchmark problems}

To evaluate the operation of the algorithm written in the Python, test runs were carried out to find global and local minimums on well-known test functions. Despite the fact that all the functions below are generalized for the multidimensional case, in this work we will run our program only for the two-dimensional case, since the algorithm itself can easily be rewritten for any dimensions, and in the two-dimensional case it is easier to clearly demonstrate the operation of the algorithm itself.

\begin{itemize}
\item \textit{Shekel function}. Multidimensional, multimodal, continuous, deterministic function commonly used as a test function for testing optimization techniques \citep{shekel1971test}.
\item \textit{Rosenbrock function}. Non-convex function, introduced by Howard H. Rosenbrock in 1960 \cite{rosenbrock1960automatic}, which is used as a performance test problem for optimization algorithms. The global minimum is inside a long, narrow, parabolic shaped flat valley. To find the valley is trivial. To converge to the global minimum, however, is difficult.
\item \textit{Himmelblau function}. Multi-modal function, used to test the performance of optimization algorithms. The locations of all the minima can be found analytically. However, because they are roots of cubic polynomials, when written in terms of radicals, the expressions are somewhat complicated. The function is named after David Mautner Himmelblau, who introduced it \cite{himmelblau1972applied}.
\item \textit{Rastrigin function}. Non-convex, non-linear multimodal function. It was first proposed by Rastrigin \cite{rastrigin1974extremal} as a 2-dimensional function and has been generalized by Rudolph \cite{rudolph1990globale}. The generalized version was popularized by Hoffmeister and B\"ack \cite{hoffmeister1990genetic} and M\"uhlenbein et al \cite{muhlenbein1991parallel}. Finding the minimum of this function is a fairly difficult problem due to its large search space and its large number of local minima.
\end{itemize}

\begin{figure}[h]
\begin{minipage}[t]{0.5\linewidth}
\center{\includegraphics[width=1\linewidth]{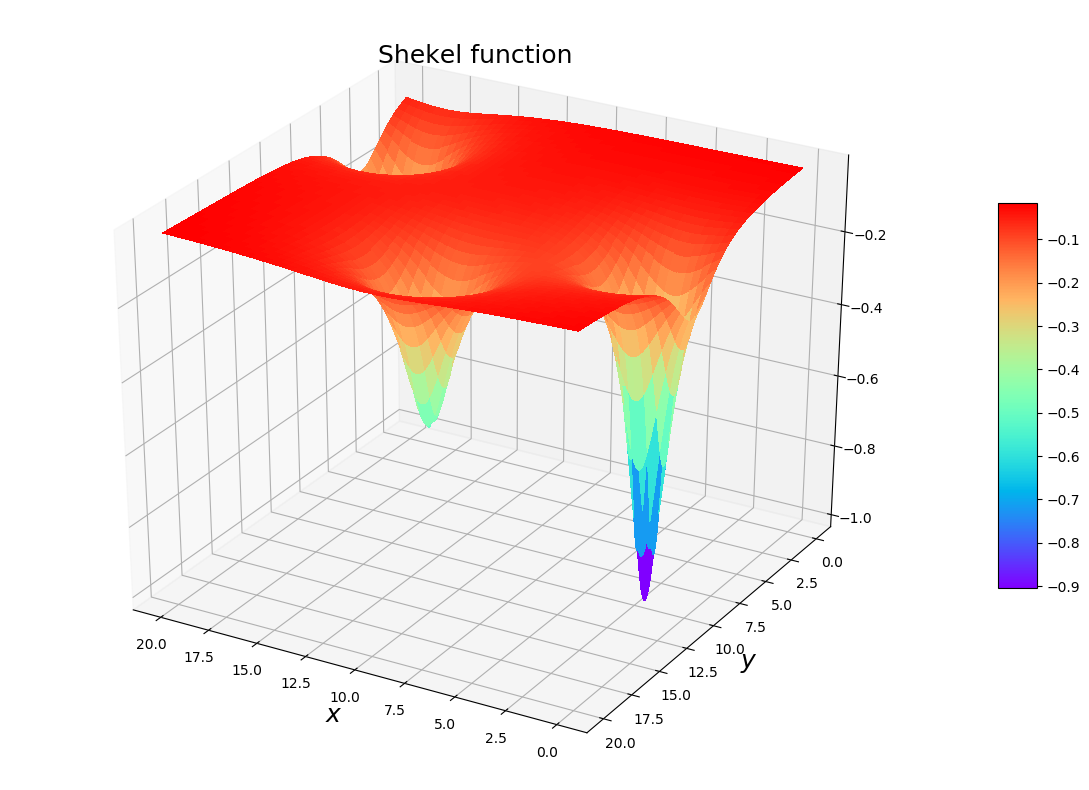}}
\caption{3D plot of Shekel function}
\label{shekel_3d}
\end{minipage}
\hfill
\begin{minipage}[t]{0.5\linewidth}
\center{\includegraphics[width=1\linewidth]{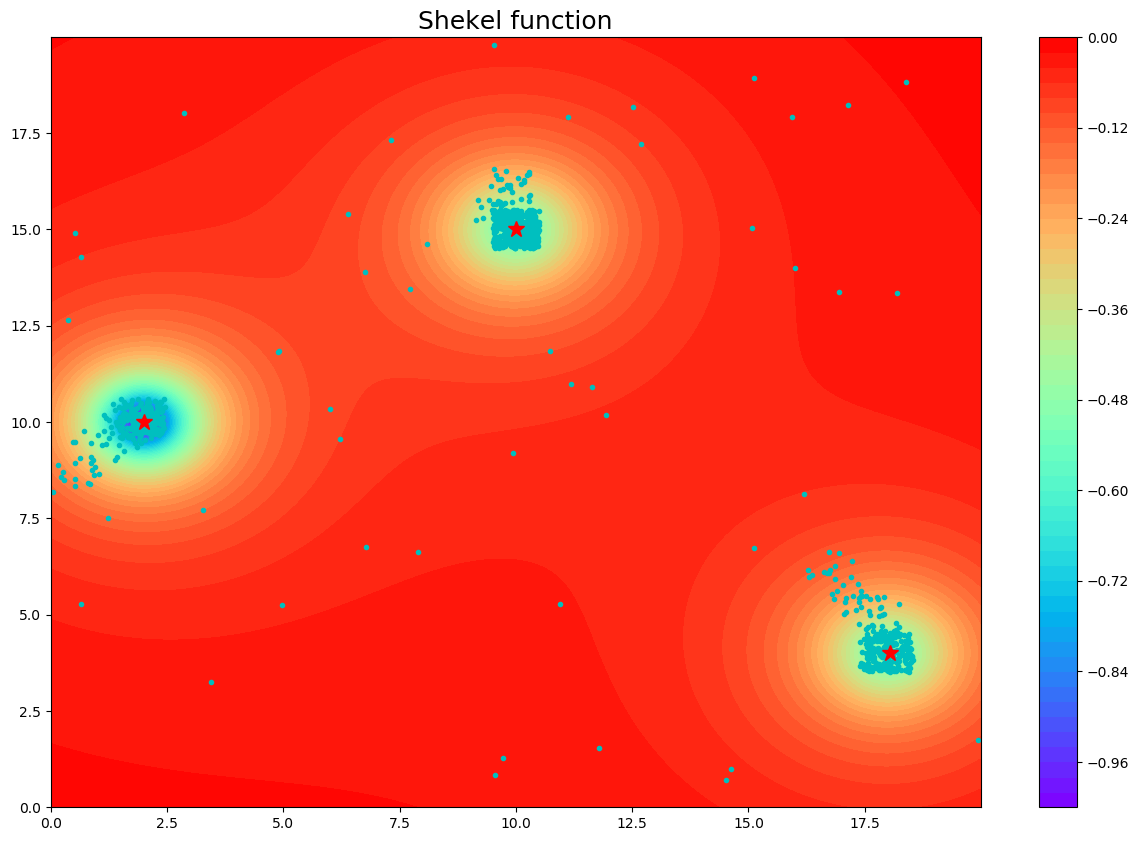}}
\caption{Result of ABC algorithm}
\label{bee_shekel}
\end{minipage}
\end{figure}

\begin{figure}[h]
\begin{minipage}[t]{0.5\linewidth}
\center{\includegraphics[width=1\linewidth]{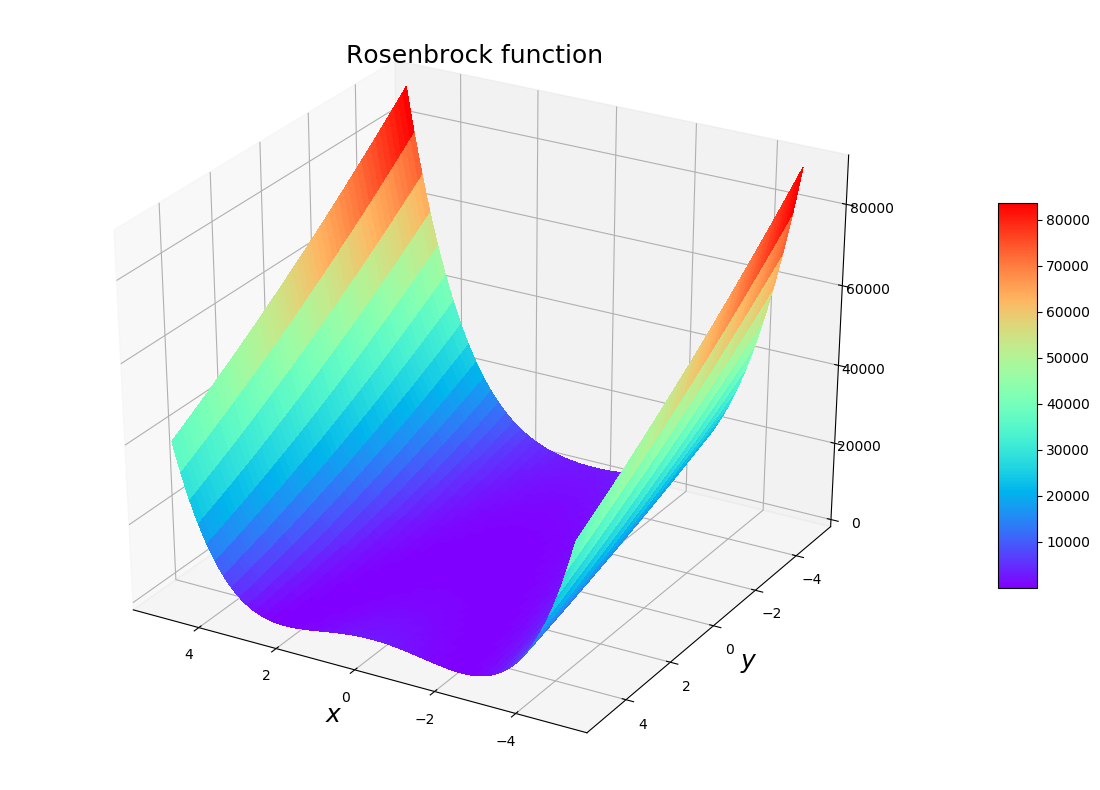}}
\caption{3D plot of Rosenbrock function}
\label{rosenbrock_3d}
\end{minipage}
\hfill
\begin{minipage}[t]{0.5\linewidth}
\center{\includegraphics[width=1\linewidth]{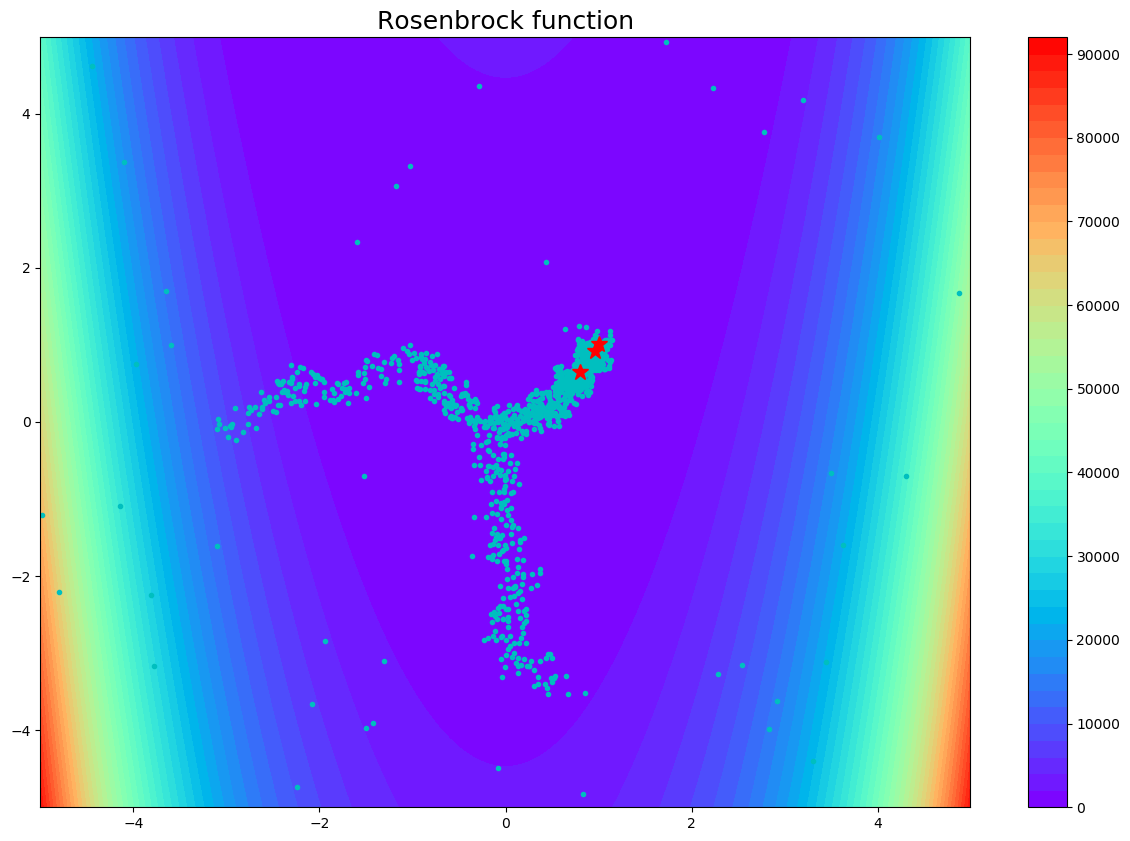}}
\caption{Result of ABC algorithm}
\label{bee_rosenbrock}
\end{minipage}
\end{figure}

\begin{figure}[h]
\begin{minipage}[t]{0.5\linewidth}
\center{\includegraphics[width=1\linewidth]{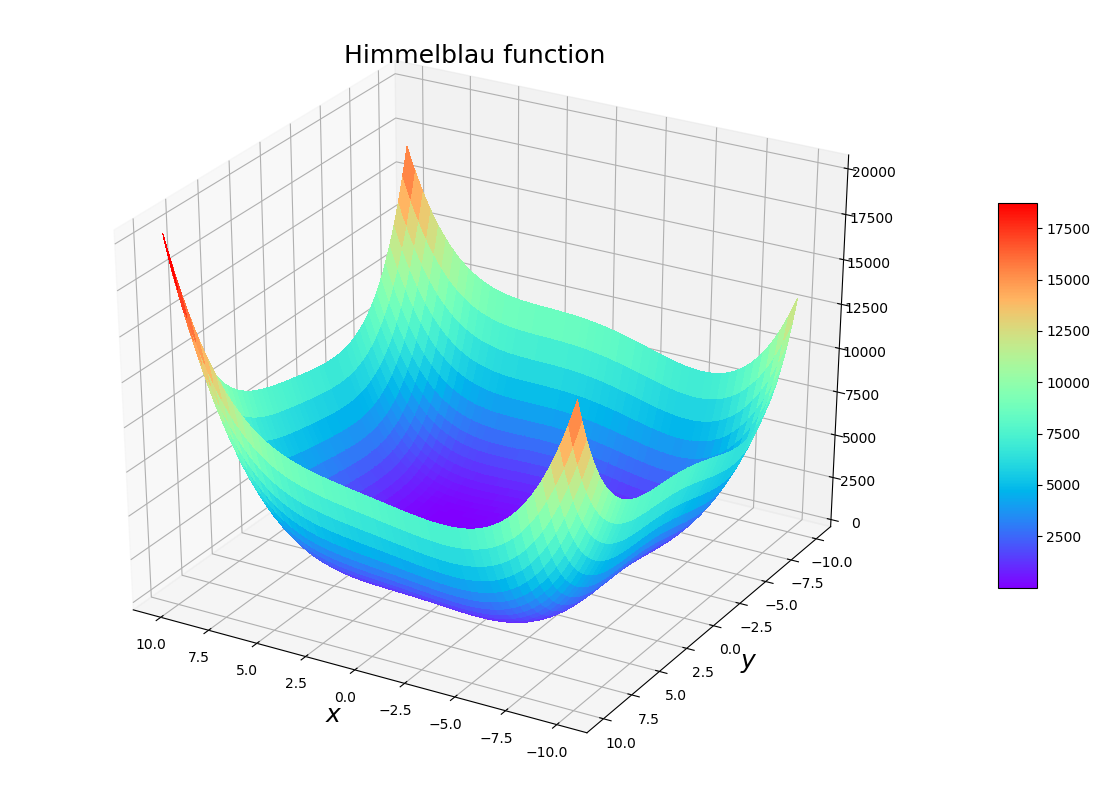}}
\caption{3D plot of Himmelblau function}
\label{himmelblau_3d}
\end{minipage}
\hfill
\begin{minipage}[t]{0.5\linewidth}
\center{\includegraphics[width=1\linewidth]{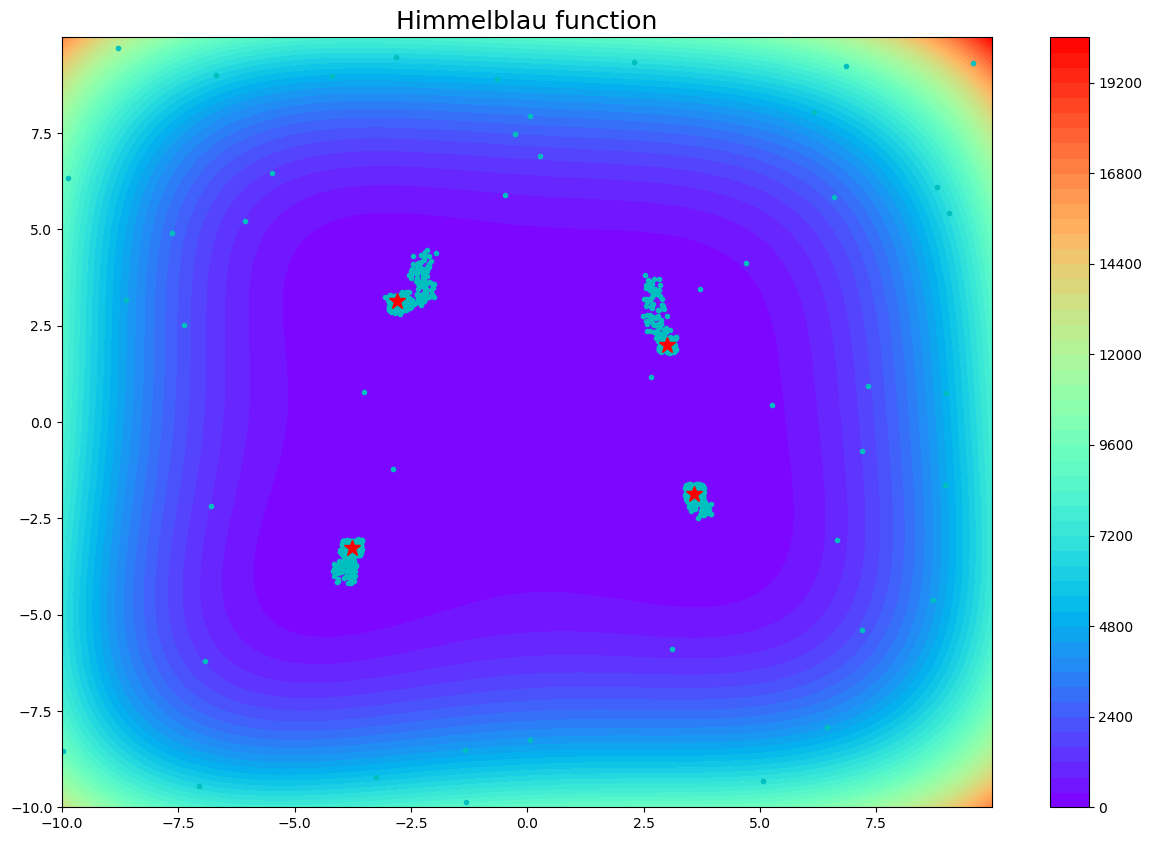}}
\caption{Result of ABC algorithm}
\label{bee_himmelblau}
\end{minipage}
\end{figure}

\begin{figure}[h]
\begin{minipage}[t]{0.5\linewidth}
\center{\includegraphics[width=1\linewidth]{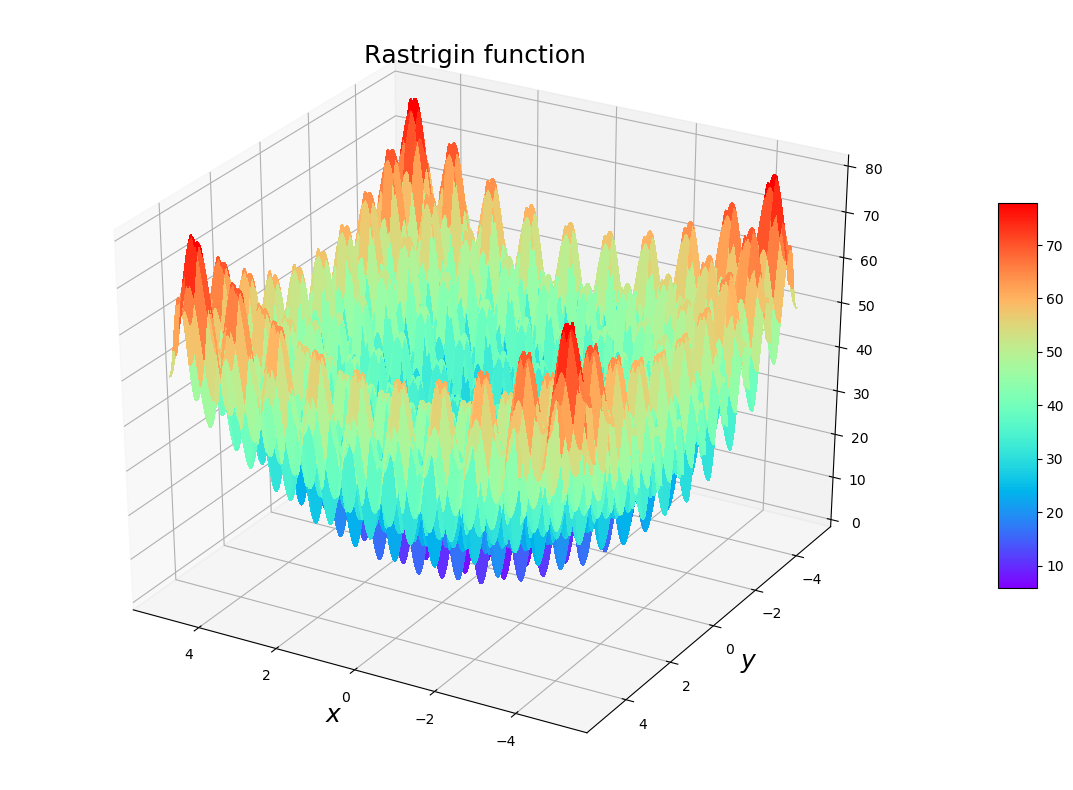}}
\caption{3D plot of Rastrigin function}
\label{rastrigin_3d}
\end{minipage}
\hfill
\begin{minipage}[t]{0.5\linewidth}
\center{\includegraphics[width=1\linewidth]{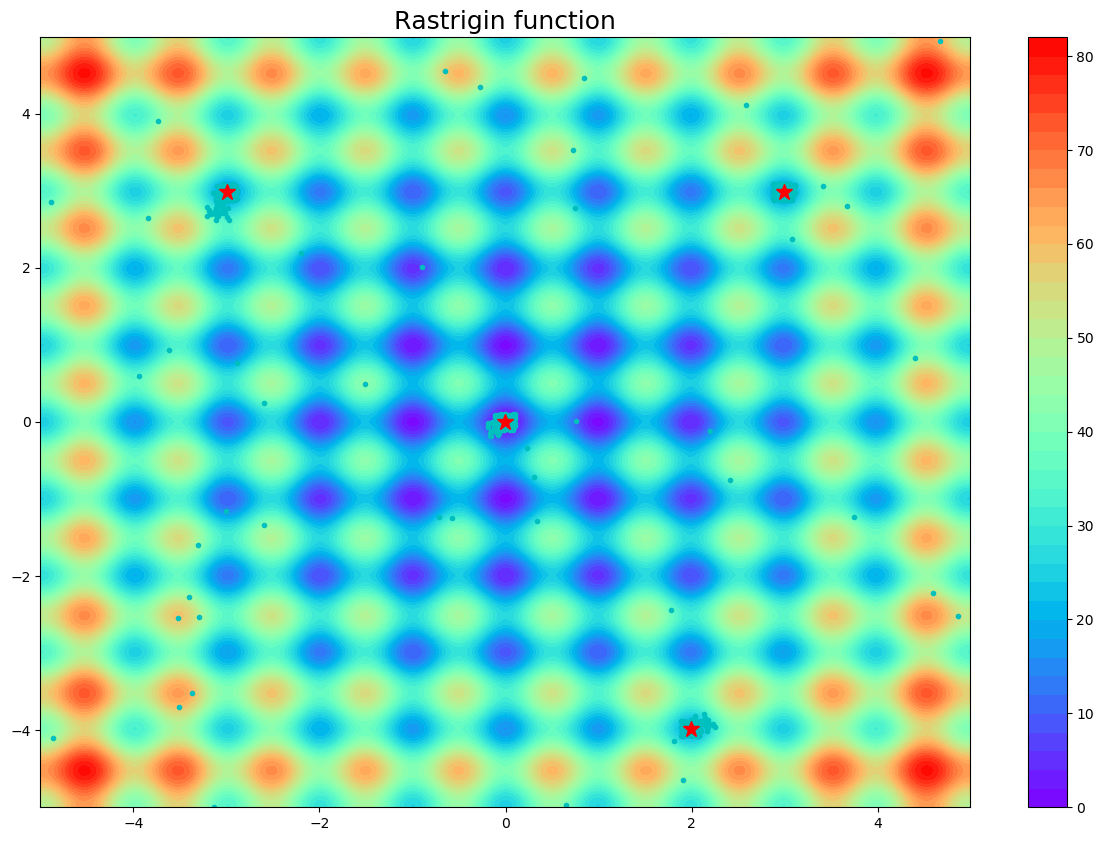}}
\caption{Result of ABC algorithm}
\label{bee_rastrigin}
\end{minipage}
\end{figure}

\FloatBarrier

\begin{sidewaystable}[]
\centering
\caption{Benchmark definitions}
\begin{tabular}{|l|l|l|l|l|}
\hline
\textbf{Function}   & \textbf{Formulae} & \textbf{Domain} \\ \hline
Shekel     &  $F(x,y) = -\frac{1}{1+(x-2)^2 + (y-10)^2} - \frac{1}{2+(x-10)^2 + (y-15)^2} - \frac{1}{2+(x-18)^2 + (y-4)^2}$  &  $D = \{(x,y) | x \in [0, 20], y \in [0, 20]\}$ \\ \hline
Rosenbrock &  $F(x, y) = 100(y - x^2)^2 + (1-x)^2$  & $D = \{(x,y) | x \in [-5, 5], y \in [-5, 5]\}$ \\ \hline
Himmelblau & $F(x,y) = (x^2 + y - 11)^2 + (x + y^2 - 7)^2$ & $D = \{(x,y) | x \in [-10, 10], y \in [-10, 10]\}$  \\ \hline
Rastrigin  & $F(x,y) = 20 + x^2 + y^2 - 10(cos(2\pi x) + cos(2\pi y))$ & $D = \{(x,y) | x \in [-5, 5], y \in [-5, 5]\}$ \\ \hline
\end{tabular}
\label{tab_benchmarks_def}

\bigskip

\centering
\caption{Exact and found minimums}
\begin{tabular}{|l|l|l|c|c|}
\hline
\textbf{Function} & \textbf{Minimum}                                                                                                           & \textbf{Program output}                                                                                                                                                               & \multicolumn{1}{l|}{\textbf{Iterations}} & \multicolumn{1}{l|}{\textbf{NFE}} \\ \hline
Shekel  & \begin{tabular}[c]{@{}l@{}}F(2, 10) = -1.01439037\\ F(10, 15) = -0.5165\\ F(18, 4) = -0.5088\end{tabular}                                    & \begin{tabular}[c]{@{}l@{}}F(2.01238, 10.00536) = -1.01424\\ F(9.99079, 14.9984) = -0.51645\\ F(18.00263, 4.00842) = -0.50875\end{tabular}                                   & 93                              & 1260                     \\ \hline
Rosenbrock & F(1, 1) = 0                                                                                                                                  & \begin{tabular}[c]{@{}l@{}}F(1.00434, 1.00954) =  9.134e-05\\ F(0.96131, 0.92385) =  0.0015\\ F(0.80291, 0.64176) =  0.03969\end{tabular}                                    & 138                             & 1630                     \\ \hline
Himmelblau & \begin{tabular}[c]{@{}l@{}}F(3.584428, -1.848126) = 0\\ F(-2.805118, -3.131312) = 0\\ F(-3.779310, -3.283186) = 0\\ F(3, 2) = 0\end{tabular} & \begin{tabular}[c]{@{}l@{}}F(3.57899, -1.8563) = 0.00284\\ F(-2.80447, 3.1313) = 1.356e-05\\ F(-3.77731, -3.27165) = 0.00543\\ F(3.00457, 2.00039) = 0.00081\end{tabular}    & 113                             & 1650                     \\ \hline
Rastrigin  & F(0,0) = 0                                                                                                                                   & \begin{tabular}[c]{@{}l@{}}F(-0.00168, -0.00483) = 0.00518\\ F(-2.98775, 2.98489) = 17.91085\\ F(2.98575, 2.97744) = 17.92021\\ F(1.99025, -3.97938) = 19.89912\end{tabular} & 96                              & 1470                     \\ \hline
\end{tabular}
\label{tab_minimums}
\end{sidewaystable}

\FloatBarrier

The definitions for each function you can see in table \ref{tab_benchmarks_def}. The 3D plots of each functions you can see in Figs. \ref{shekel_3d}, \ref{rosenbrock_3d}, \ref{himmelblau_3d}, \ref{rastrigin_3d}. The operation of the algorithm can be seen in the Figs. \ref{bee_shekel}, \ref{bee_rosenbrock}, \ref{bee_himmelblau}, \ref{bee_rastrigin}. The \textit{light blue} spots means processed points, \textit{red stars} indicate the extremum found in each locations. The functions of Shekel and Himmelblau are quite simple in terms of finding extrema, so the number of the best locations and perspective locations in total were chosen exactly to the number of extrema. Since the functions of Rosenbrock and Rastrigin are rather complex functions, despite the fact that they have only one global minimum, we gave the algorithm to work out more locations to increase probability of success. Also, it is extremely important for us to control the Number of Function Evaluations (NFE), because for our main problem, which is described in the next section, computing many direct problems is expensive. The list of minimums and program results can be found in table \ref{tab_minimums}. As you can see, the algorithm works fine in all benchmarks. All plots in this article built using the matplotlib library \cite{Hunter2007}.

\section{Mathematical model}
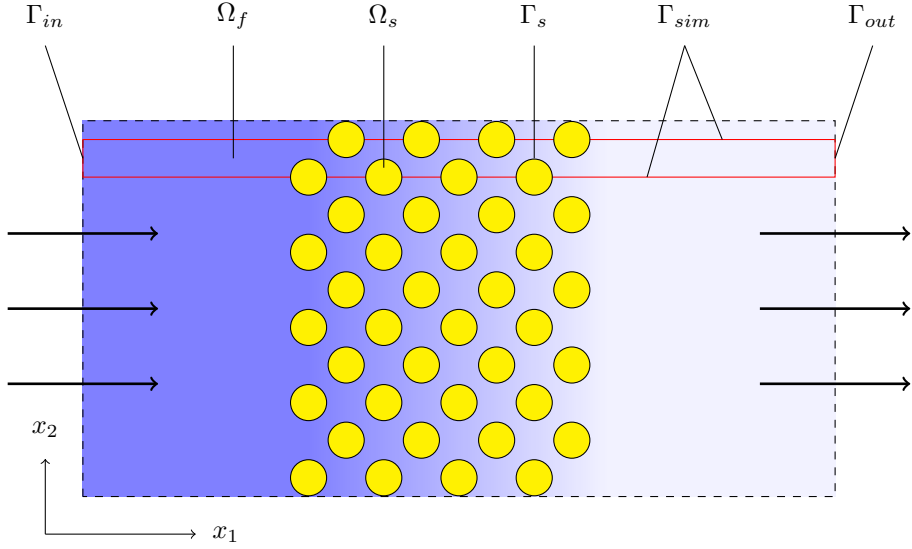
\begin{figure}[ht] 
  \begin{center}
    \begin{tikzpicture}
       \shade[top color=blue!5, bottom color=blue!5] (7,0) rectangle +(3,5);
       \shade[left color=blue!50, right color=blue!5] (3,0) rectangle +(4,5);
       \shade[top color=blue!50, bottom color=blue!50] (0,0) rectangle +(3,5);
       \draw [dashed] (0, 0) rectangle +(10,5);
       \draw [color=red] (0,4.25) rectangle +(10,0.5);
       \foreach \y in {3,...,6} 
	\foreach \x in {0,...,4} {
  	\filldraw [fill=yellow,draw=black] (\y,\x+0.25) circle (0.24);
  	\filldraw [fill=yellow,draw=black] (\y+0.5,\x+0.75) circle (0.24);
       }
       \draw [<-, line width=1, color=black] (1,1.5) -- (-1,1.5);
       \draw [<-, line width=1, color=black] (1,2.5) -- (-1,2.5);
       \draw [<-, line width=1, color=black] (1,3.5) -- (-1,3.5);
       \draw [<-, line width=1, color=black] (11,1.5) -- (9,1.5);
       \draw [<-, line width=1, color=black] (11,2.5) -- (9,2.5);
       \draw [<-, line width=1, color=black] (11,3.5) -- (9,3.5);
       \draw [-] (2,4.5) -- (2,6);
       \draw  (2,6.4) node {$\Omega_f$};
       \draw [-] (4,4.375) -- (4,6);
       \draw  (4,6.4) node {$\Omega_s$};  
       \draw [-] (6,4.5) -- (6,6);
       \draw  (6,6.4) node {$\Gamma_s$}; 
       \draw [-] (0,4.5) -- (-0.5,6);
       \draw  (-0.5,6.4) node {$\Gamma_{in}$};
       \draw [-] (7.5,4.25) -- (8,6);
       \draw [-] (8.5,4.75) -- (8,6);
       \draw  (8,6.4) node {$\Gamma_{sim}$};
       \draw [-] (10,4.5) -- (10.5,6);
       \draw  (10.5,6.4) node {$\Gamma_{out}$};  
       \draw [->] (-0.5,-0.5) -- (1.5,-0.5);
       \draw  (1.9,-0.5) node {$x_1$};  
       \draw [->] (-0.5,-0.5) -- (-0.5,0.5);
       \draw  (-0.5,0.9) node {$x_2$};  
    \end{tikzpicture}
    \caption{Sketch of the pore scale domain} 
   \label{f-1}
  \end{center}
\end{figure}

In this work, we will use absolutely the same mathematical model which described in our previous work \cite{grigoriev2019computational}, where we investigate diffusion dominated case, where our Damk\"oler numbers are small. At this work, we will deal with reaction dominated case, where we will have big Damk\"oler numbers and in this situation, finding a minimum can be difficult. For simplicity of analysis we will consider the porous media as periodically arranged circles in two dimensional case. Scheme of the domain is shown on Fig.\ref{f-1}. Due to the periodicity, we will only generate a computational grid from a segment of the entire domain which marked with a red rectangle on Fig.\ref{f-1}. A part of the domain, namely $\Omega_f$ is occupied by a fluid, while the other part is occupied be the obstacles $\Omega_s$. The obstacle surfaces (where the reaction occurs) are denoted by $\Gamma_s$, while symmetry lines are denoted by $\Gamma_{sim}$. It is supposed that dissolved substance is introduced via the inlet boundary $\Gamma_{in}$, and the part of the substance which did not react flows out via $\Gamma_{out}$. Since the adsorption reaction occurs only on the surface of obstacles, the computational domain consists only of $\Omega_f$. 

\subsection{Flow problem} 

The flow in the pores described here by the steady state incompressible Stokes equations:
\begin{equation}\label{1}
\nabla p - \mu \nabla^2 \bm{u} = 0,
\end{equation} 
\begin{equation}\label{2}
 \nabla \cdot \bm{u} = 0, 
 \quad  \bm{x} \in \Omega_f, 
 \quad  t > 0,
\end{equation} 
where $\bm{u}(\bm{x})$ and $p(\bm{x})$ are the fluid velocity and pressure, respectively,
while $\mu > 0$ and $\rho > 0$ are the viscosity and the density, which we assume to be constants \cite{bear2013dynamics,acheson2005elementary}. 


Denote by $\bm n$ the outer normal vector to the boundary. 
Suitable boundary conditions on $\partial \Omega_f$ are specified. 
The velocity of the fluid $\bar{u}$ is prescribed at the inlet 
\begin{equation}\label{3}
 \bm{u} \cdot \bm n = \bar{u},
 \quad \bm{u} \times  \bm n = 0,
 \quad \bm x \in \Gamma_{in} .
\end{equation} 
At the outlet, pressure and absence of tangential force are prescribed 
\begin{equation}\label{4}
 p - \bm \sigma \bm n \cdot \bm n = \bar{p},
 \quad \bm \sigma \bm n \times \bm n = 0, 
 \quad \bm x \in \Gamma_{out} . 
\end{equation}  
Standard no-slip and no-penetration conditions are prescribed on the solid walls:
\begin{equation}\label{5}
 \bm{u} \cdot \bm n = 0,
 \quad \bm{u} \times  \bm n = 0,
 \quad \bm x \in \Gamma_{s} .
\end{equation}
Symmetry conditions are prescribed on the symmetry boundary of the computational domain:
\begin{equation}\label{6}
 \bm{u} \cdot \bm n = 0,
 \quad \bm \sigma \bm n \times \bm n = 0, 
 \quad \bm x \in \Gamma_{sim} .  
\end{equation} 

\subsection{Species Transport} 

The concentration of the solute in the fluid is denoted by $c(\bm x, t)$. The unsteady solute transport in absence of homogeneous reactions is governed by convection diffusion equation
\begin{equation}\label{7}
 \frac{\partial c }{\partial t} + \nabla (\bm{u} c) 
 - D \nabla^2 c = 0 ,
 \quad  \bm{x} \in \Omega_f, 
 \quad  t > 0,
\end{equation} 
where $D > 0$ is the solute diffusion coefficient which is assumed to be scalar and constant. 

The concentration of the solute at the inlet is assumed to be known:
\begin{equation}\label{8}
 c(\bm x, t) = \bar{c},
 \quad \bm x \in \Gamma_{in} ,
\end{equation} 
where $\bar{c} > 0$ is assumed to be constant.
Zero diffusive flux of the solute at the outlet and on
the external boundaries of the domain is prescribed as follows:
\begin{equation}\label{9}
 D  \nabla c \cdot \bm{n} = 0,
 \quad \bm x \in \Gamma_{sim} \cup  \Gamma_{out} .
\end{equation} 
Note that convective flux via the outlet is implicitly allowed by the above equations.
The surface reactions that occur at the obstacles' surface $\Gamma_s$ satisfy the mass conservation law, in this particular case meaning that the change in adsorbed surface concentration is equal to the flux from the fluid to the surface. This is described as
\begin{equation}\label{10}
 \frac{\partial m}{\partial t} = - D  \nabla c \cdot \bm{n} ,
 \quad \bm x \in \Gamma_{s},
\end{equation} 
where $m$ is the surface concentration of adsorbed solute \cite{kralchevsky2008chemical}.
A mixed kinetic–diffusion adsorption description is used:
\begin{equation}\label{11}
 \frac{\partial m}{\partial t} = f(c, m) .
\end{equation} 
For reactive boundaries, the choice of $f$  and its dependence on $c$ and $m$ is critical for a correct description of the reaction dynamics at the solid–fluid interface. A number of different isotherms (i.e., different functions $f(c,m)$) exist for describing these dynamics, dependent on the solute attributes, the order of the reaction, and the interface type.

The simplest of these is the Henry isotherm, which assumes a linear relationship between the near surface concentration and the surface concentration of the adsorbed particles, and takes the form
\begin{equation}\label{12}
 f(c, m) = k_a c - k_d m,
\end{equation} 
Here $k_a \geq 0$ is the rate of adsorption, measured in unit length per unit time, 
and $k_d \geq 0$ is the rate of desorption, measured per unit time. This isotherm we used in our previous work in diffusion dominated case investigation \cite{grigoriev2019computational}. In this work we use the Langmuir adsorption isotherm which is more complicated, three parameter model:
\begin{equation}\label{13}
 f(c, m) = k_a c \left (1 - \frac{m}{m_\infty} \right )  - k_d m .
\end{equation} 
Here $m_\infty > 0$  is the maximal possible adsorbed surface concentration. 
In comparison to the Henry isotherm (\ref{12}), the Langmuir isotherm (\ref{13}) predicts a decrease in the rate of adsorption as the adsorbed concentration increases due to the reduction in available adsorption surface. 

The formulation of the initial -- boundary value problem in addition to the governing equations  
(\ref{1}), (\ref{2}), (\ref{7}), boundary conditions (\ref{3})--(\ref{6}), (\ref{8})--(\ref{11}), and specified isotherm (\ref{12}) or (\ref{13}), requires specification of initial conditions:
\[
 \bm{u}(\bm x, 0) = \bm{\bar{u}}(\bm x), 
 \quad \bm x \in \Omega_f , 
\]
\begin{equation}\label{14}
 c(\bm x, 0) = c_0(\bm x), 
 \quad \bm x \in \Omega_f . 
\end{equation} 
\begin{equation}\label{15}
 m(\bm x, 0) = m_0(\bm x), 
 \quad \bm x \in \Gamma_s. 
\end{equation} 


\subsection{Dimensionless form of the equations} 

When a problem like the above one need to be solved for a range of parameters (what is our goal here),
working with dimensionless form of the equations give definitive advantages.
For the dimensionless variables (velocity, pressure, concentration) below, the same notations are used as for the dimensional ones. The height of the computational domain $\Omega_f$, namely $l$, is used for scaling spatial sizes, the scaling of the velocity is done by the inlet velocity $\bar{u}$, and the scaling of the concentration is done by the inlet concentration $\bar{c}$.

The Stokes Eq.(\ref{3}) and its boundary conditions in dimensionless remain unchanged, keeping mind that in this case they are written with respect to dimensionless velocity and pressure, and considering the dimensionless viscosity to be equal to one.

In dimensionless form Eq. (\ref{7}) reads
\begin{equation}\label{20}
 \frac{\partial c }{\partial t} + \nabla (\bm{u} c) 
 - \frac{1}{\mathrm{Pe}}  \nabla^2 c = 0 ,
 \quad  \bm{x} \in \Omega_f, 
 \quad  t > 0,
\end{equation} 
where
\[
 \mathrm{Pe} =  \frac{l \bar{u} }{D} 
\] 
is the Peclet number.

Further on, Eq. (\ref{8}) is transformed into 
\begin{equation}\label{21}
 c = 1,
 \quad \bm x \in \Gamma_{in} ,
\end{equation} 
while the boundary condition (\ref{9}) take the form
\begin{equation}\label{22}
 \nabla c \cdot \bm{n} = 0,
 \quad \bm x \in \Gamma_{sim} \cup  \Gamma_{out} .
\end{equation} 
The dimensionless form of Eq.(\ref{10}) is given by
\begin{equation}\label{23}
 \frac{\partial m}{\partial t} = - \nabla c \cdot \bm{n} ,
 \quad \bm x \in \Gamma_{s},
\end{equation}
where $m$ is scaled as follows:
\[
 \bar{m} =  l \bar{c} .
\]  
In dimensionless form the adsorption relations, in the case of Henry isotherm, are written as follows
\begin{equation}\label{24}
 \frac{\partial m}{\partial t} = \mathrm{Da}_a c - \mathrm{Da}_d m,
 \quad \bm x \in \Gamma_{s},
\end{equation}
where the adsorption and desorption Damkoler numbers are given by
\[
 \mathrm{Da}_a = \frac{k_a}{\bar{u}} ,
 \quad  \mathrm{Da}_d = \frac{k_d l}{\bar{u}} .
\] 
In the case when we consider Langmuir isotherm, (\ref{11}), (\ref{13}), the following dimensionless relation is used
\begin{equation}\label{25}
 \frac{\partial m}{\partial t} = \mathrm{Da}_a c \left (1- \frac{m}{\mathrm{M}} \right )   - \mathrm{Da}_d m,
 \quad \bm x \in \Gamma_{s},
\end{equation}
where the dimensionless parameter $M$ is given by 
\[
 \mathrm{M} = \frac{m_\infty }{l \bar{c}} .
\] 


\section{Numerical solution of the direct problem} 

Finite Element Method, FEM, is used for space discretization of the above problem, together with implicit discretization in time. The algorithm used here for solving the direct problem is practically identical of the algorithm used to study oxidation in \cite{oxidation}.

\subsection{Geometry and grid} 

The computational domain is a rectangle with a dimensionless height of $x_2=1$ and dimensionless length of $x_1=17.5$, in which ten half cylinders are embedded.
The distance between the centers of the cylinders in $x_1$ direction is 1.5 dimensionless units, the radius of cylinders is 0.4 dimensionless units.

The computational domain $\Omega_f$ is triangulated using the grid generator Gmsh (website gmsh.info) \cite{Gmsh}. 
The script for preparing the geometry is written in Python. A demonstration of the computational grid is shown in Fig.{\ref{mesh_grid}.

\begin{figure}[h!]
\center{\includegraphics[width=1\linewidth]{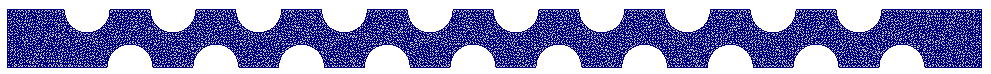}}
\caption{Computational mesh.}
\label{mesh_grid}
\end{figure}
\FloatBarrier

\subsection{Computation of steady state single phase fluid flow} 

One way coupling is considered here. The fluid flow influences the species transport, but there is no back influence of the species concentration on the fluid flow. Based on this, the flow is computed in advance. The FEM approximation of the steady state flow problem \cite{gresho200incompressible} is based on variational formulation of the considered boundary value problem (\ref{1}), (\ref{2}), (\ref{3}), (\ref{4}), (\ref{5}), (\ref{6}). 
The following functional space $\bm V$ is defined for the velocity $\bm u$  ($\bm u \in \bm V$):
\[
\begin{split}
 \bm V = \{ \bm v \in \bm H^1(\Omega) : \  & \bm{u} \cdot \bm n = 1, \
 \bm{u} \times  \bm n = 0 \ \mathrm{on} \ \Gamma_{in} , 
 \\ & \bm{u} = 0 \ \mathrm{on} \ \Gamma_{s}, \ \bm{u} \cdot \bm n = 0 \
 \ \mathrm{on} \ \Gamma_{sim} \} .
\end{split} 
\] 
Test function $\bm v \in \hat{\bm V}$, where
\[
 \hat{\bm V} = \{ \bm v \in \bm H^1(\Omega_f) : \ \bm{u} = 0 \ \mathrm{on} \ \Gamma_{in} , \ 
 \bm{u} = 0 \ \mathrm{on} \ \Gamma_{s}, \ \bm{u} \cdot \bm n = 0 \
 \ \mathrm{on} \ \Gamma_{sim} \} .
\] 
For the pressure $p$ and the related test functions $q$, it is required that $p, q \in Q$, where 
\[
 Q = \{ g \in L_2(\Omega_f) : \ q = 0 \ \mathrm{on} \  \Gamma_{out} \} .
\] 

Let us multiply Eq.(\ref{1}) by $\bm v$, Eq.(\ref{2}) by $q$, and integrate over the computational domain. Taking into account the boundary conditions (\ref{3}), (\ref{4}), (\ref{5}), (\ref{6}), the following system of equations is obtained
with respect to $\bm v \in \bm V$, $q \in Q$
\begin{equation}\label{26}
 a(\bm u, \bm v) - b(\bm v, p) = 0 \ \forall \bm v \in \hat{\bm V} ,
\end{equation} 
\begin{equation}\label{27}
 b(\bm u, q) = 0 \ \forall q \in Q .
\end{equation} 
Here
\[
  a(\bm u, \bm v) := \int_{\Omega_f} \nabla \bm u \cdot \nabla \bm v \, d \bm x , 
\] 
\[
 b(\bm v, p) := \int_{\Omega_f}(\nabla \cdot \bm v) q \, d \bm x . 
\] 

For the FEM approximation of the velocity, the pressure, and the respective test functions, the following finite dimensional subspaces are selected
$\bm V_h \subset \bm V$, $\hat{\bm V}_h \subset \hat{\bm V}$
and $Q_h \subset Q$.
Taylor-Hood $P_2-P_1$ elements \cite{taylor1973numerical} are used here.
These are continuous $P_2$ Lagrange elements for the velocity components 
and continuous $P_1$ Lagrange elements for the pressure field.
The computations are carried out using the computing platform for partial differential equations FEniCS (website fenicsproject.org) \cite{LoggMardalEtAl2012a,AlnaesBlechta2015a}.

\begin{figure}[h!]
\center{\includegraphics[width=1\linewidth]{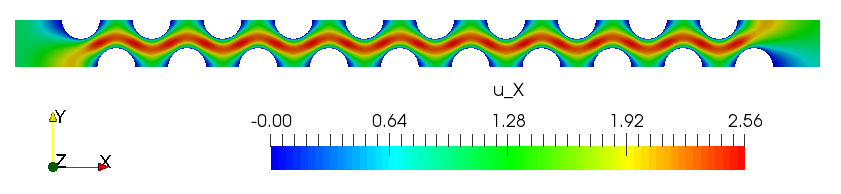}}
\caption{Velocity field by $X$ direction.}
\label{u_x}
\end{figure}

\begin{figure}[h!]
\center{\includegraphics[width=1\linewidth]{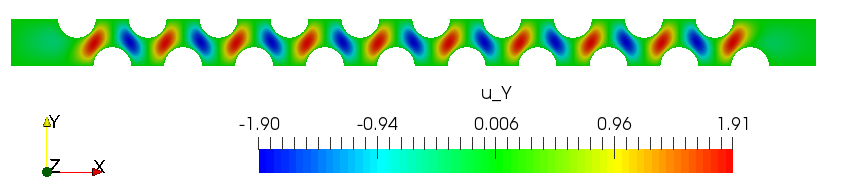}}
\caption{Velocity field by $Y$ direction.}
\label{u_y}
\end{figure}

\begin{figure}[h!]
\center{\includegraphics[width=1\linewidth]{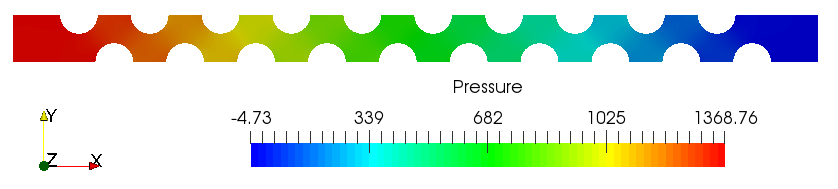}}
\caption{Pressure field.}
\label{press}
\end{figure}
\FloatBarrier

As mentioned above, mainly slow flows are of interest for the current study, therefore the basic considered variant is characterized by $\mathrm{Re} =1$. Convergence of the solution with respect to refinement of the grid is illustrated on Fig.\ref{spatial}.  Convergence of the solution with respect to time step is illustrated on Fig.\ref{temporal}.
We have used three computational grids: basic grid with 18743 nodes and 35958 triangles, coarse grid with 4760  nodes and 8754 triangles and fine grid with 72745  nodes and 142460 triangles.
From the results, it can be concluded that the coarse grid provides good enough accuracy for the numerical solution. Taking into account that we will solve a lot of direct problems, it was decided to conduct research on a coarse grid.

Before starting work, it was decided to conduct research on the parameters, namely:\\
\begin{itemize}
\item $\mathrm{Pe} = [1, 10, 100]$
\item $\mathrm{Da}_a = [50, 100, 200]$
\item $\mathrm{Da}_d = [0.5, 1, 2]$
\item $\mathrm{M}_\infty = [100, 1000, 100000]$
\end{itemize}
The results illustrated on Figs.\ref{pe_diffs},\ref{m_diffs},\ref{daa_diffs},\ref{dad_diffs}. The curve in these figures is a breakthrough curve which describes how much solute has crossed the outflow boundary and it is given by
\begin{equation}
c_{out}(t) = \frac{\int_{\Gamma_{out}}c(\bm{x}, t)d\bm{x}}{\int_{\Gamma_{out}}d\bm{x}}.
\end{equation}

\begin{figure}[h]
\begin{minipage}[t]{0.47\linewidth}
\center{\includegraphics[width=1\linewidth]{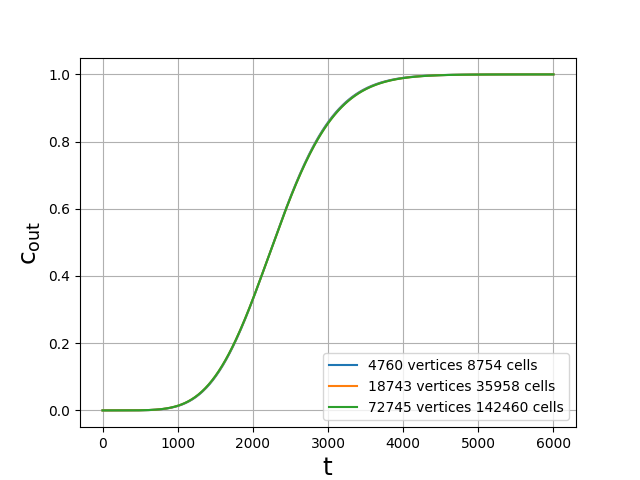}}
\caption{Research on computational grids of different densities.}
\label{spatial}
\end{minipage}
\hfill
\begin{minipage}[t]{0.47\linewidth}
\center{\includegraphics[width=1\linewidth]{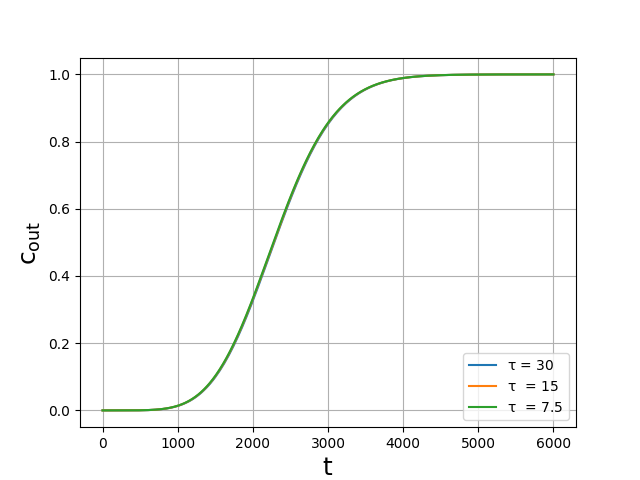}}
\caption{Research on different time steps.}
\label{temporal}
\end{minipage}
\end{figure}

\begin{figure}[h]
\begin{minipage}[t]{0.47\linewidth}
\center{\includegraphics[width=1\linewidth]{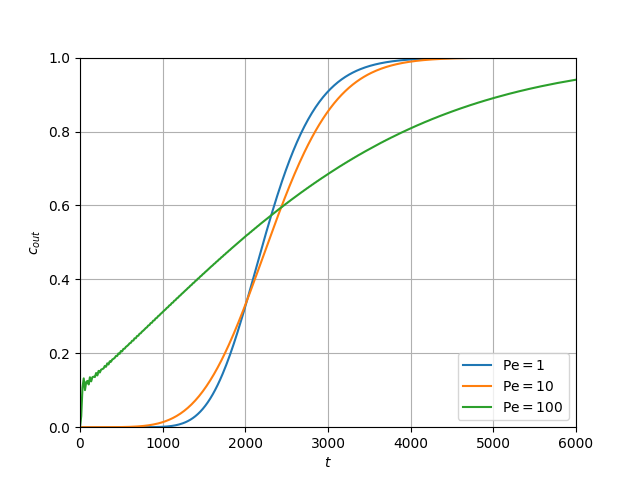}}
\caption{Research on different Peclet numbers.}
\label{pe_diffs}
\end{minipage}
\hfill
\begin{minipage}[t]{0.47\linewidth}
\center{\includegraphics[width=1\linewidth]{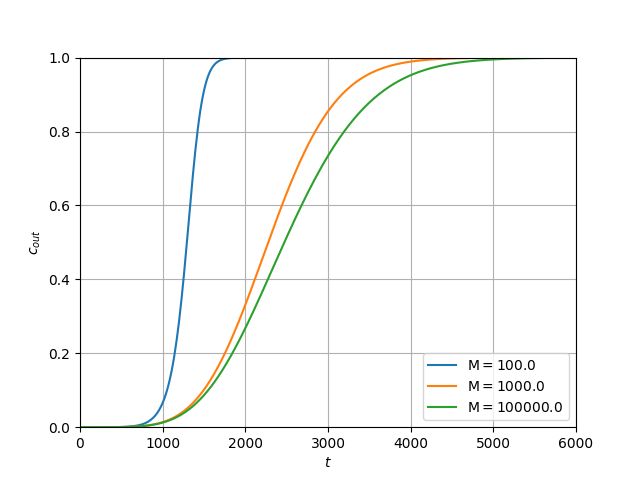}}
\caption{Study on different maximum concentrations adsorbed on the surface.}
\label{m_diffs}
\end{minipage}
\end{figure}

\begin{figure}[h]
\begin{minipage}[t]{0.47\linewidth}
\center{\includegraphics[width=1\linewidth]{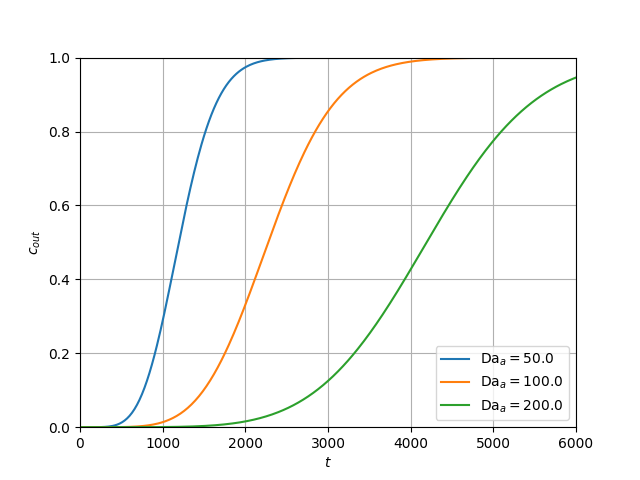}}
\caption{Research on different Damk\"oler numbers for adsorption.}
\label{daa_diffs}
\end{minipage}
\hfill
\begin{minipage}[t]{0.47\linewidth}
\center{\includegraphics[width=1\linewidth]{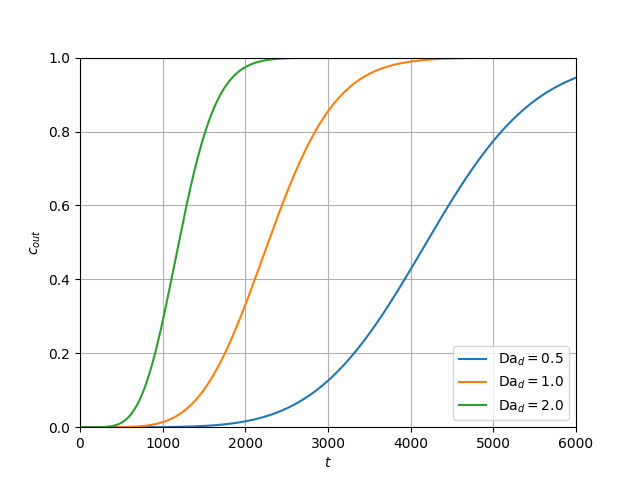}}
\caption{Research on different Damk\"oler numbers for desorption.}
\label{dad_diffs}
\end{minipage}
\end{figure}
\FloatBarrier

\subsection{Simulation of reactive transport} 

The unsteady species transport problem (\ref{20}),  (\ref{15}), (\ref{21})-- (\ref{23}) is solved numerically using standard Lagrangian $P_1$ finite elements. Let us define 
\[
 S = \{ s \in H^1(\Omega_f) : \ q = 1 \ \mathrm{on} \  \Gamma_{in} \} ,
\]  
\[
 \hat{S} = \{ s \in H^1(\Omega_f) : \ q = 0 \ \mathrm{on} \  \Gamma_{in} \} .
\] 
The approximate solution $c \in S$ is sought from
\begin{equation}\label{28}
\left (\frac{\partial c}{\partial t}, s \right) + d(c,s) =  
 \left (\frac{\partial m}{\partial t} ,s \right )_s
 \quad \forall s \in \hat{S} ,   
\end{equation} 
where the following notations are used
\[
  d(c,s) := - \int_{\Omega_f} c \bm u \cdot \nabla  s \, d \bm x 
  + \frac{1}{\mathrm{Pe}} \int_{\Omega_f} \nabla c \cdot \nabla  s \, d \bm x 
  + \int_{\Gamma_{out}} (\bm u \cdot \bm n) c s \, d \bm x ,
\] 
\[
  (\varphi,s)_s := - \int_{\Gamma_{s}} \varphi   s \, d \bm x .
\] 
For determining $m \in G = L_2(\Gamma_{s})$ (see (\ref{11})) we use
\begin{equation}\label{29}
 \left (\frac{\partial m}{\partial t} , g \right )_s -
 (f(c,m), g)_s = 0,
 \quad g \in G .  
\end{equation} 
The discretization in time is based on symmetric discretization (Crank–Nicolson method), which is second order accurate (see, e.g., \cite{Samarskii,Ascher2008}. 
Let $\tau$  be a step-size of a uniform grid in time such that 
$c^n = c(t^n), t^n = n\tau, \ n = 0, 1, ...$. 

Eq.(\ref{28}) is approximated in time as follows
\[
\left (\frac{c^{n+1} - c^n}{\tau }, s \right) + d\left (\frac{c^{n+1} + c^n}{2} ,s\right ) =  
 \left (\frac{m^{n+1} - m^n}{\tau } ,s \right )_s .
\] 
Similarly, for (\ref{29}) we get
\[
\left (\frac{m^{n+1} - m^n}{\tau }  , g \right )_s -
 \left (f\left (\frac{c^{n+1} + c^n}{2} ,\frac{m^{n+1} + m^n}{2} \right ), g\right )_s = 0 ,
 \quad n = 0, 1, ... , 
\] 

In the considered here case, zero initial conditions are posed
\[
 c^0 = 0, \quad \bm x \in \Omega_f, 
\] 
\[
 m^0 = 0, \quad \bm x \in \Gamma_s . 
\]

Species concentration at different time moments are shown on Fig.\ref{diff_time_solute}. 

Sensitivity study is carried out in order to see how the change of different parameters leads to change of the breakthrough curves. Such sensitivity studies are often first stage for optimization or parameter identification procedures. The dependence of the average outlet concentration from Peclet number is shown on Fig.\ref{pe_diffs}.

The rate of adsorption is characterized by $\mathrm{Da}_a$. The influence of this parameter on the outlet concentration is illustrated on Fig.\ref{daa_diffs}. Increasing $\mathrm{Da}_a$, as expected, leads to more intensive adsorption and larger amount of the deposited substance. The influence of the parameter $\mathrm{Da}_d$ on the outflow concentration is illustrated on Fig.\ref{dad_diffs}.

Henry isotherm usually describes well the initial stages of the adsorption. In cases when only a limited mass can be adsorbed at a surface, Langmuir isotherm should be used to reflect the decay of the adsorption rate close to the saturation. In this case an additional parameter appears, namely $\mathrm{M}$ (see (\ref{25})).
The influence of $\mathrm{M}$ on the average output concentration is shown on Fig.\ref{m_diffs}.

\begin{figure}[h]
\begin{minipage}[t]{1.0\linewidth}
\center{\includegraphics[width=1\linewidth]{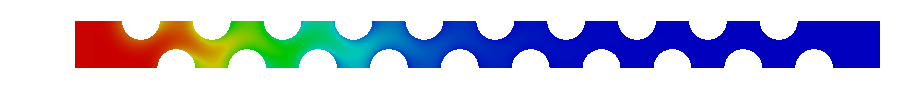}} \\a
\label{t_450}
\end{minipage}
\hfill
\begin{minipage}[t]{1.0\linewidth}
\center{\includegraphics[width=1\linewidth]{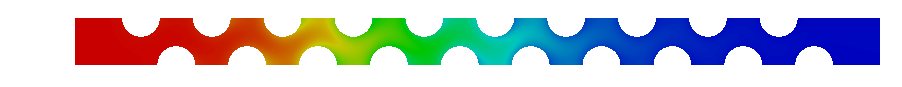}} \\b
\label{t_900}
\end{minipage}
\hfill
\begin{minipage}[t]{1.0\linewidth}
\center{\includegraphics[width=1\linewidth]{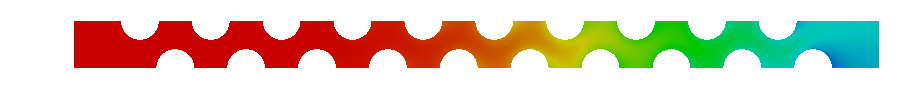}} \\c
\label{t_1800}
\end{minipage}
\hfill
\begin{minipage}[t]{1.0\linewidth}
\center{\includegraphics[width=1\linewidth]{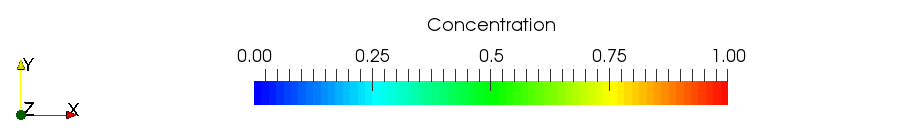}}
\label{conc_shk}
\end{minipage}
\caption{Species concentration at different dimensionless time moments: a --- $t = 450$ , b --- $t = 900$, c --- $t = 1800$}
\label{diff_time_solute}
\end{figure}
\FloatBarrier

\section{Numerical solution of the inverse problem in reaction dominated case} 

Consider an inverse problem for determining unknown adsorption rate (so called parameter identification problem). As a function which we want to minimize, we choose the functional of the residual which is given by
\begin{equation}
J(\mathrm{Da}_a, \mathrm{Da}_d, \mathrm{M}_{\infty}) = \int_0^T (c_{out}(t) - \widehat{c(t)} )^2 dt,
\end{equation}
where $\widehat{c(t)}$ is a breakthrough curve from experimental set. The starting point is monitoring of the difference (residual) between the measured $\widehat{c_{out}(t)}$ and the computed  $c_{out}(t)$ average outflow concentration for different values of the parameters $\mathrm{Da}_a$,  $\mathrm{Da}_d$ and $\mathrm{M}_\infty$ in Langmuir isotherm Eq.(\ref{25}).  

\subsection{Deterministic approach}
Taking into account all the results shown in the figures above, it was decided to use a grid with 4760 nodes and the following parameters: $\tau = 30, \mathrm{Pe} = 10, \mathrm{Da}_a = 100, \mathrm{Da}_d = 1, \mathrm{M}_\infty = 1000$. To evaluate the residual functional, we decided to fix each of the three parameters and solve the functional values on a uniform grid of 50 by 50 (see results on Figs. \ref{fixed_daa}, \ref{fixed_dad}, \ref{fixed_m}). \textit{Blue star} is our exact parameters location, \textit{red contour} is an admissible set. As you can see, our functional has a rather complex shape. In the case when we do not know the location of exact parameters, in 3D case we must build uniform grid, for example 50 by 50 by 50, which is equal to 125000 direct calculations of the problem. Obviously, this strategy is very effective, simple, it is impossible to make a mistake with it, but almost always it is a quite expensive for calculation. We made this investigations only to see the form of our functional, because, first of all, this work is devoted to the operation of the bee colony algorithm.

\begin{figure}[h]
\begin{minipage}[t]{0.45\linewidth}
\center{\includegraphics[width=1\linewidth]{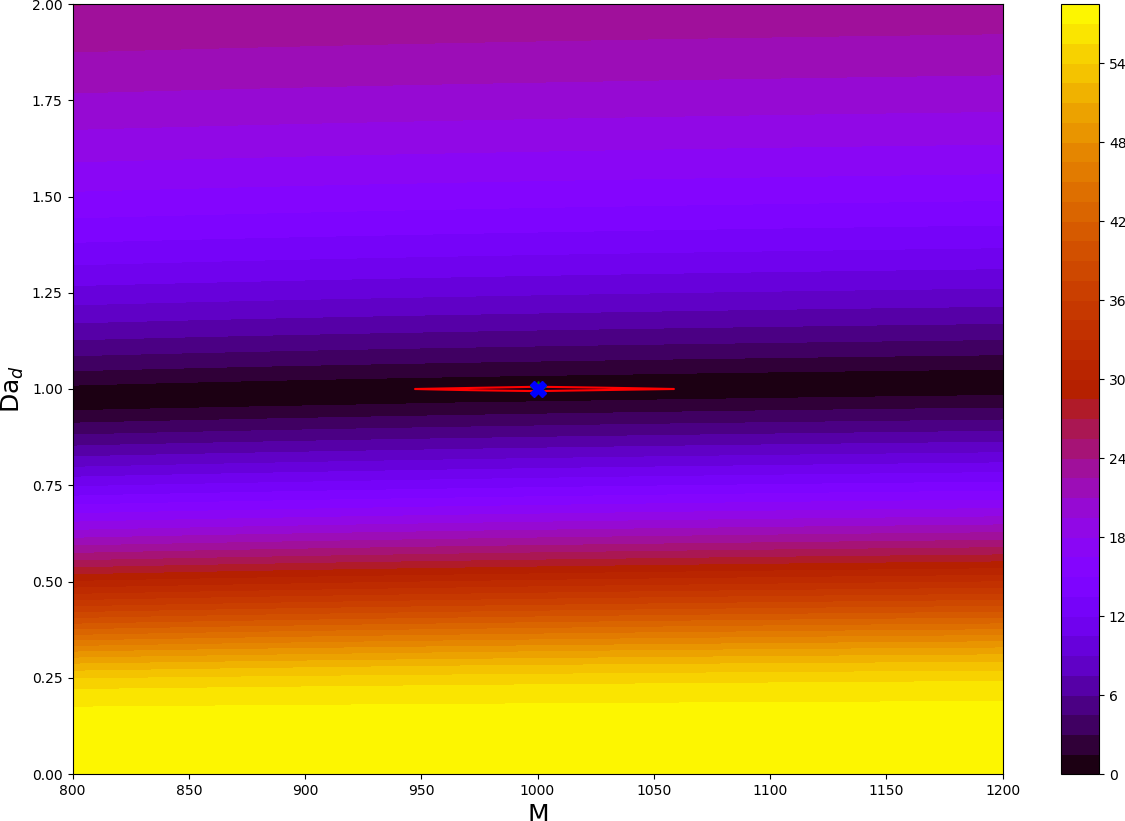}}
\end{minipage}
\hfill
\begin{minipage}[t]{0.45\linewidth}
\center{\includegraphics[width=1\linewidth]{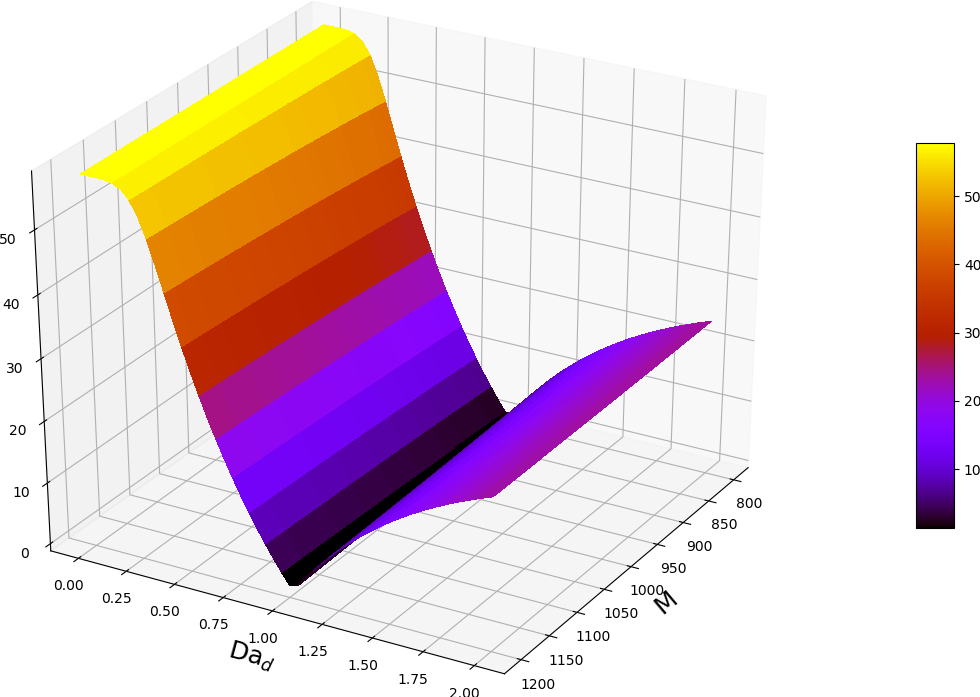}}
\end{minipage}
\caption{Calculations for fixed $\rm{Da_a}$}
\label{fixed_daa}
\end{figure}

\begin{figure}[h]
\begin{minipage}[t]{0.45\linewidth}
\center{\includegraphics[width=1\linewidth]{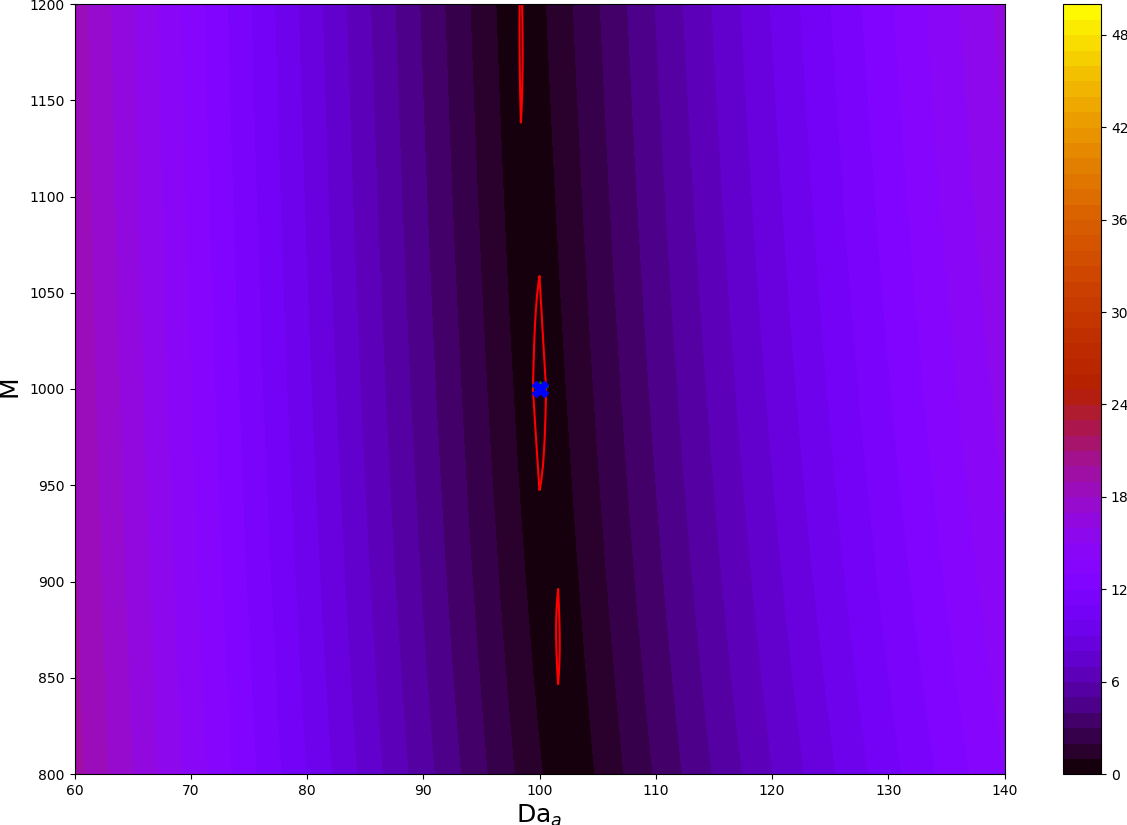}}
\end{minipage}
\hfill
\begin{minipage}[t]{0.45\linewidth}
\center{\includegraphics[width=1\linewidth]{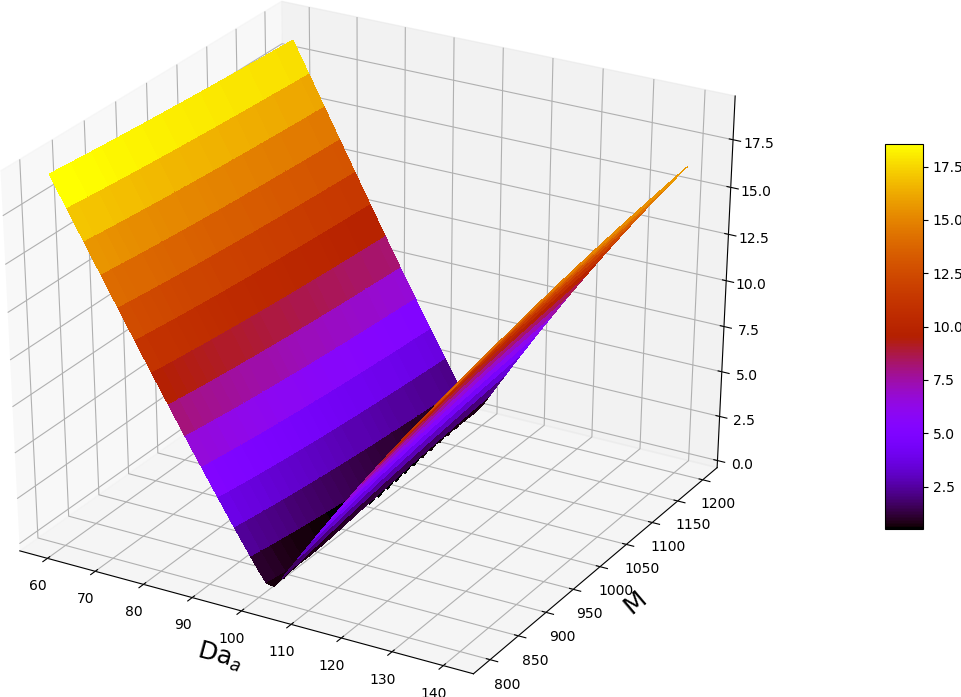}}
\end{minipage}
\caption{Calculations for fixed $\rm{Da_d}$}
\label{fixed_dad}
\end{figure}

\begin{figure}[h]
\begin{minipage}[t]{0.45\linewidth}
\center{\includegraphics[width=1\linewidth]{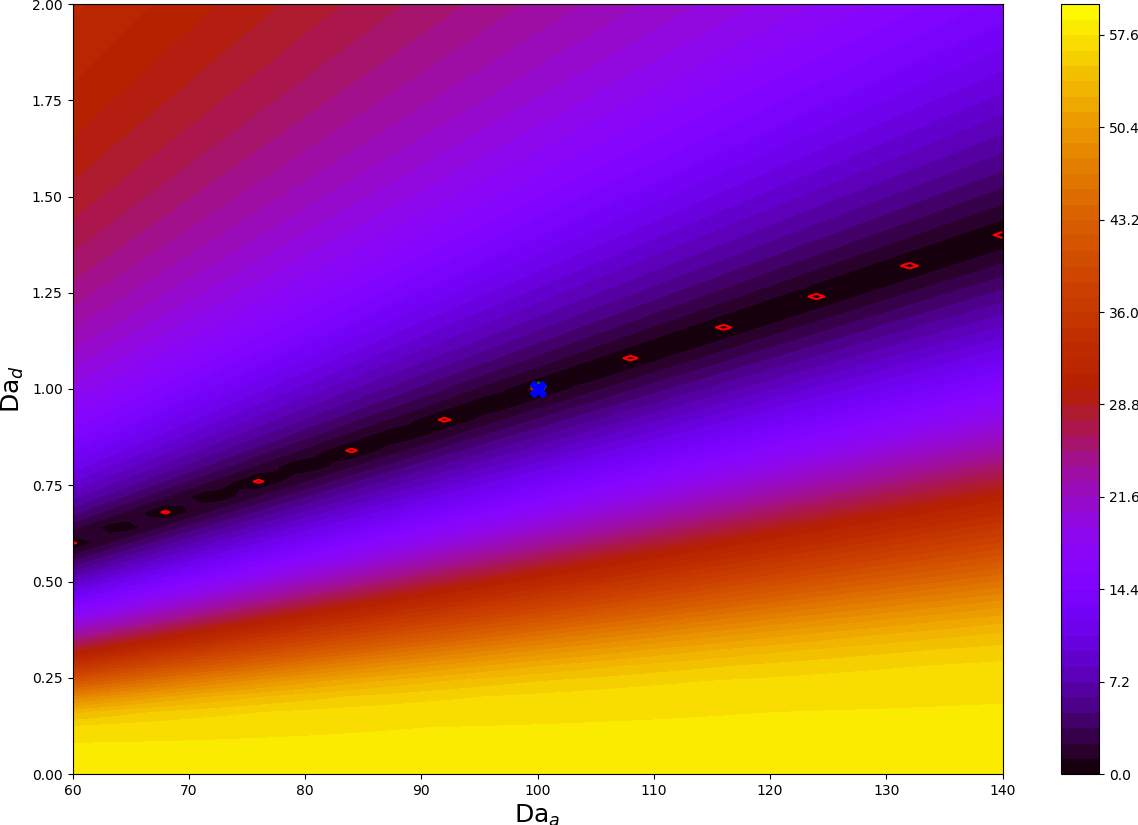}}
\end{minipage}
\hfill
\begin{minipage}[t]{0.45\linewidth}
\center{\includegraphics[width=1\linewidth]{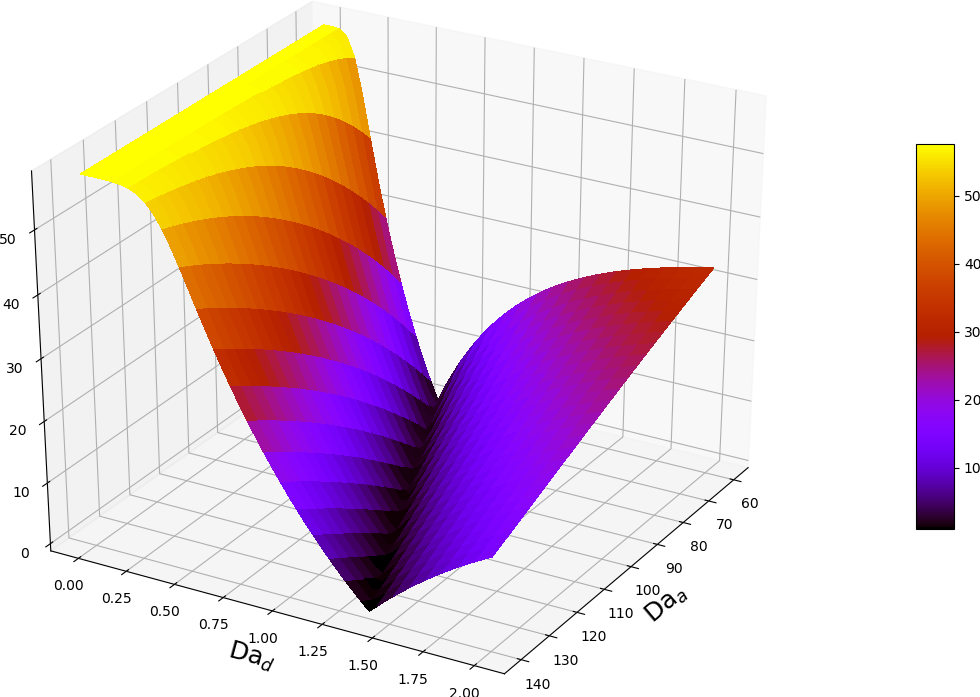}}
\end{minipage}
\caption{Calculations for fixed $\rm{M}$}
\label{fixed_m}
\end{figure}
\FloatBarrier

\subsection{Modified Bee Colony Algorithm}

Let's take the parameters: $\mathrm{Da}_a = 100$, $\mathrm{Da}_d = 1$, $\mathrm{M}_\infty = 1000$ as the exact solution that we will try to identify. We will assume that the Peclet number is known and equal to 10. Specifically, to show the behavior of the algorithm with various input parameters, we will run the algorithm two times with different sets of parameters: 
\begin{itemize}
\item[I:]  $n = 1$, $m = 2$, $d_x = 1.5$, $d_y = 0.015$, $d_z = 5$, $\delta = 1$, $sb = 20$, $abb = 20$, $abp = 5$, $\tau = 3$;
\item[II:] $n = 2$, $m = 5$, $d_x = 1.5$, $d_y = 0.015$, $d_z = 5$, $\delta = 1$, $sb = 200$, $abb = 50$, $abp = 40$, $\tau = 5$.
\end{itemize}

$X$ axis --- parameter $\mathrm{Da}_a = [60, 140]$, $Y$ axis --- parameter $\mathrm{Da}_d = [0, 2]$, $Z$ axis --- parameter $\mathrm{M}_{\infty} = [800, 1200]$.

At the end of the algorithm, the following results were obtained: \\

For parameter set I:\\
J(125.8217,1.2586,1002.1758) = 0.0087\\
J(129.8189,1.3040,1045.4760) = 0.0466\\
J(128.2248,1.2780,967.3870) = 0.0517\\
J(100.6993,1.0117,1051.8844) = 0.0615\\
Total direct problem computing (times): 2080\\

For parameter set II:\\
J(92.7936, 0.9277, 997.6068) = 0.00129\\
J(83.3579, 0.8331, 994.8119) = 0.0030\\
J(101.6545, 1.0166, 1000.0540) = 0.0042\\
J(136.4043, 1.3656, 1011.4730) = 0.0057\\
J(117.6360, 1.1779, 1014.2707) = 0.0112\\
J(108.3891, 1.0849, 1011.3793) = 0.0118\\
J(86.7377, 0.8657, 980.6640) = 0.0183\\
Total direct problem computing (times): 12300\\

The execution time of the algorithm is skipped because it depends on the computer and cost of the direct problem. As mentioned above, the number of direct problem evaluations is important to us. The quantity depends on the input parameters, so you can adjust the parameters depending on the computational cost of the direct problem and available computational power. Check Fig.\ref{MBC_res_3d} to see the results of 3D problem. One more way to decrease the number of direct problem evaluation is to increase stop criteria parameter. On all of our calculations it is set as 10e-8.

\begin{figure}[h!]
\begin{minipage}[t]{0.5\linewidth}
\center{\includegraphics[width=1.0\linewidth]{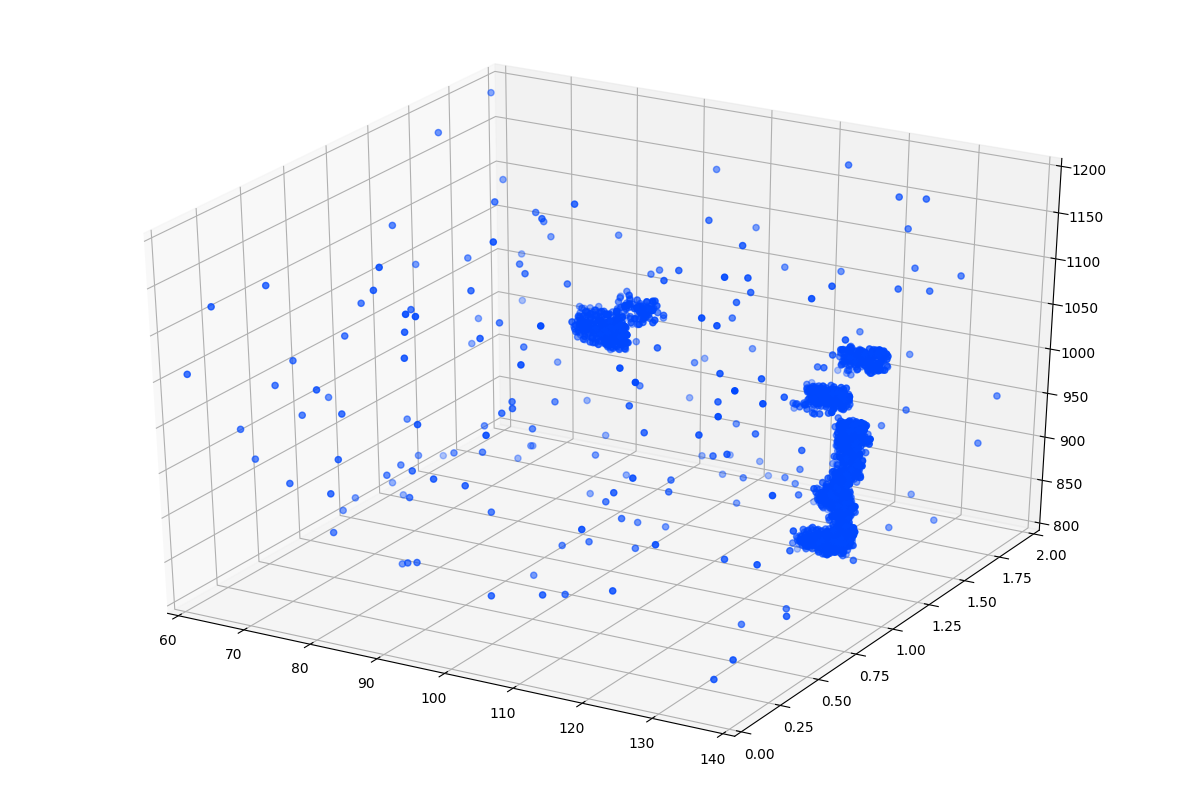}} \\ a
\end{minipage}
\hfill
\begin{minipage}[t]{0.5\linewidth}
\center{\includegraphics[width=1.0\linewidth]{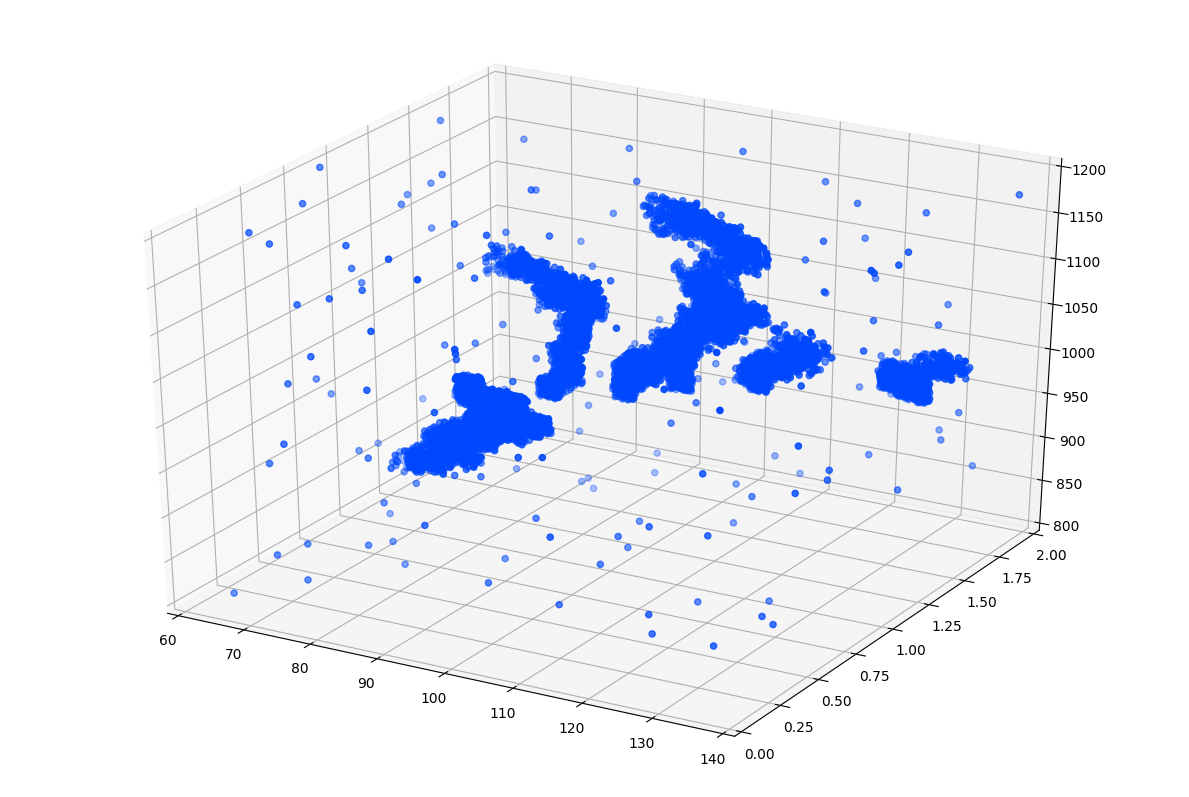}} \\ b
\end{minipage}
\caption{Results of the algorithm: a --- parameter set I, b --- parameter set II.}
\label{MBC_res_3d}
\end{figure}
\FloatBarrier

We compared the resulting extremums from both parameter sets with our exact solution and get following results: 0.0128\% and 0.0018\% of relative error for first coordinate in I and II parameter sets respectively. Of course, in some cases solving even 2080 direct problems is already expensive. So, you can increase the stop criteria parameter as noted above and you can significantly reduce number of direct problem evaluation. For example, the following parameter set
$n = 1$, $m = 2$, $d_x = 1.5$, $d_y = 0.015$, $d_z = 5$, $\delta = 1$, $sb = 20$, $abb = 20$, $abp = 5$, $\tau = 3$ was launched with stop criteria equal to 10e-3 and we get the following results:\\

\noindent J(126.1104, 1.2867, 1151.4495) = 0.3617\\
J(104.8607, 1.0383, 991.0511) = 7.5122\\
J(117.2133, 1.1463, 927.1145) = 7.5122\\
Total direct problem computing (times): 205\\

We compared the resulting extremum with the coordinates (126.1104, 1.2867, 1151.4495) with our exact solution and got a relative error in 0.626\% which is a pretty good result at a fairly low price.

The formula by which the relative error was calculated:
\begin{equation}
\mathcal{E}_{rel} = \frac{\| c_{out} - \widehat{c_{out}}\|_2}{\|\widehat{c_{out}}\|_2} = \frac{\sqrt{\sum_{i=0}^{Nt}(c_i - \widehat{c_i})^2}}{\sqrt{\sum_{i=0}^{Nt} \widehat{c_i}^2}},
\end{equation}
where $Nt$ --- the count of the time intervals, $\widehat{c_{out}}$ --- our exact solution (data from experiments). 

\subsection{Experiments with diffusion dominated case}

Nevertheless, despite the fact that the whole article is aimed at studying the reaction dominated case, we want to conduct a small analysis of the diffusion dominated case. You can read more about this case in our previous articles \cite{grigoriev2019computational,10.1007/978-3-030-41032-2_12}. Let us recall key points from that work. We investigated the diffusion dominated case (see Fig.\ref{dd_deter}), when the reaction rates were small, thus there were few time layers: $T = 40$ in dimensionless time, small time step $\tau = 0.1$. We implemented many different methods for identifying the parameters, such as deterministic approach, statistical parameter identification and multistage parameter identification.

In this section we show the results of the algorithm in diffusion dominated case and Henry isotherm. 
$X$ axis --- parameter $\mathrm{Da}_a = [0, 0.01]$, $Y$ axis --- parameter $\mathrm{Da}_d = [0, 0.1]$
Exact parameters, which we will try to identify: $\mathrm{Da}_a = 0.005$, $\mathrm{Da}_d = 0.05$.\\
The algorithm was launched with the following parameters: $n = 2$, $m = 3$, $d_x = 0.0002, d_y = 0.002$, $\delta = 0.002$, $sb = 20$, $abb = 20$, $abp = 10$, $\tau = 3$.

We got following results:\\
J(0.0050001, 0.049995) = 2.79961e-05 \\
J(0.0050081, 0.050061) = 0.000768\\
J(0.0050086, 0.049906) = 0.000947\\
J(0.0049929, 0.049853) = 0.001008\\
J(0.0050096, 0.050515) = 0.002408\\
Total direct problem computing (times): 1720\\

\begin{figure}[h!]
\begin{minipage}[t]{0.45\linewidth}
\center{\includegraphics[width=1.0\linewidth]{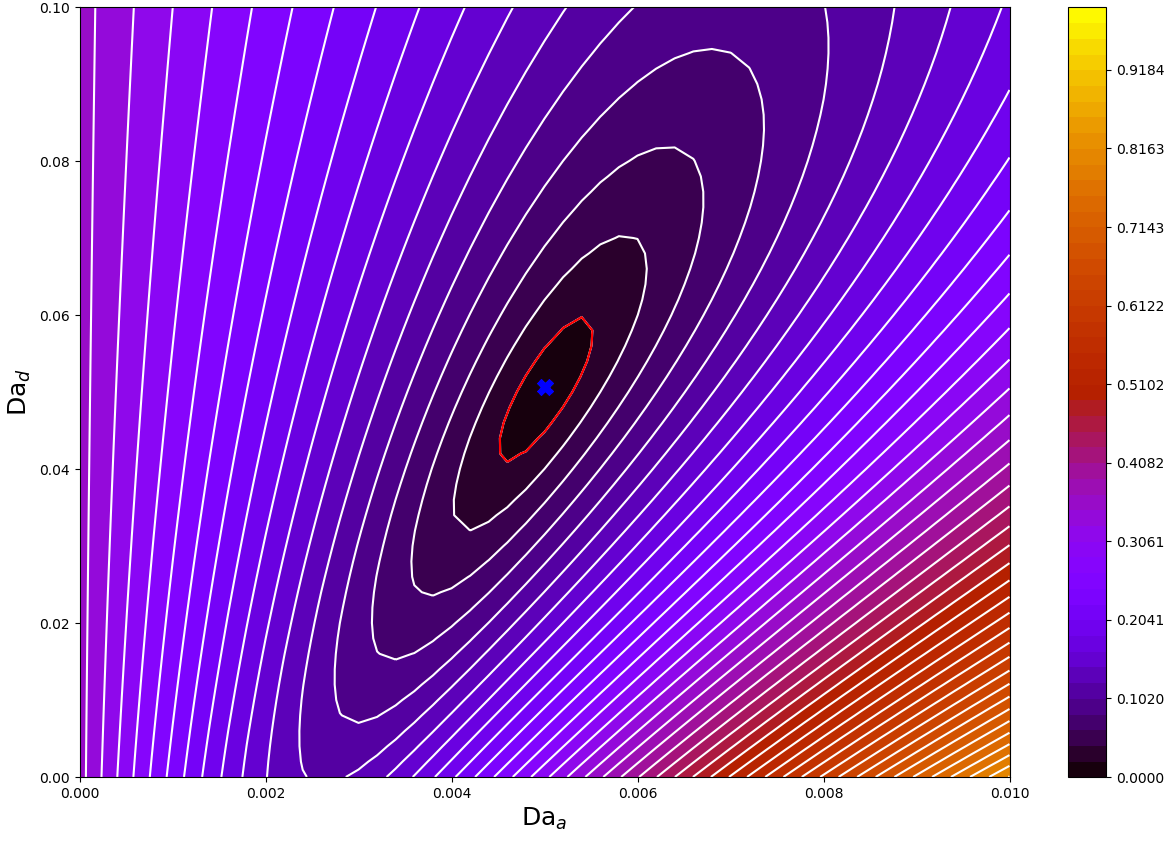}}
\end{minipage}
\hfill
\begin{minipage}[t]{0.45\linewidth}
\center{\includegraphics[width=1.0\linewidth]{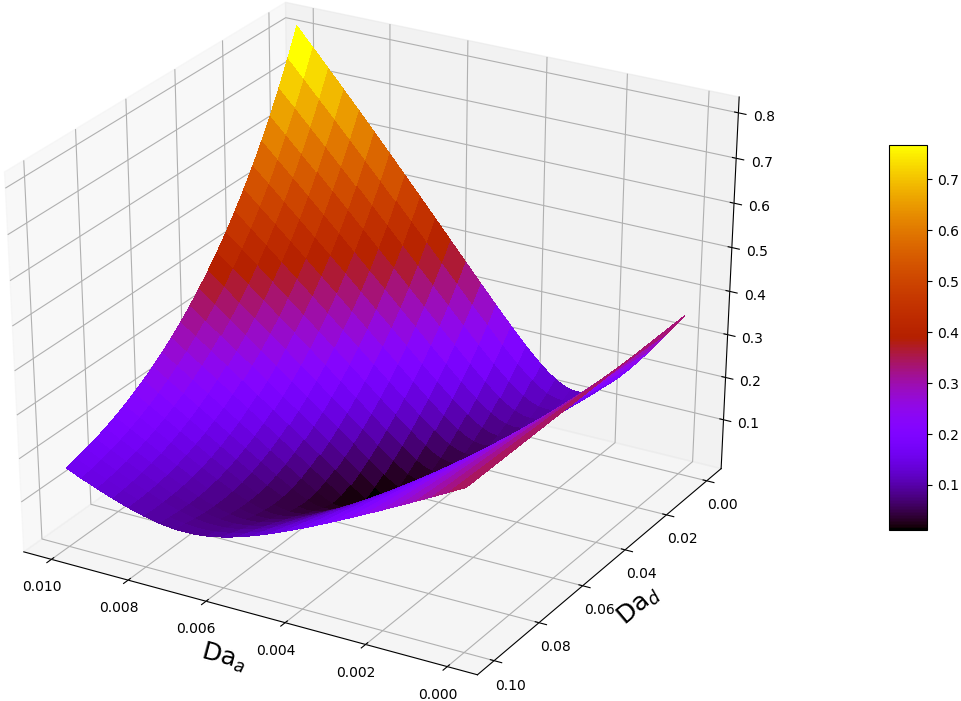}}
\end{minipage}
\caption{Diffusion dominated case}
\label{dd_deter}
\end{figure}

\begin{figure}[h!]
\begin{minipage}[t]{0.45\linewidth}
\center{\includegraphics[width=1.0\linewidth]{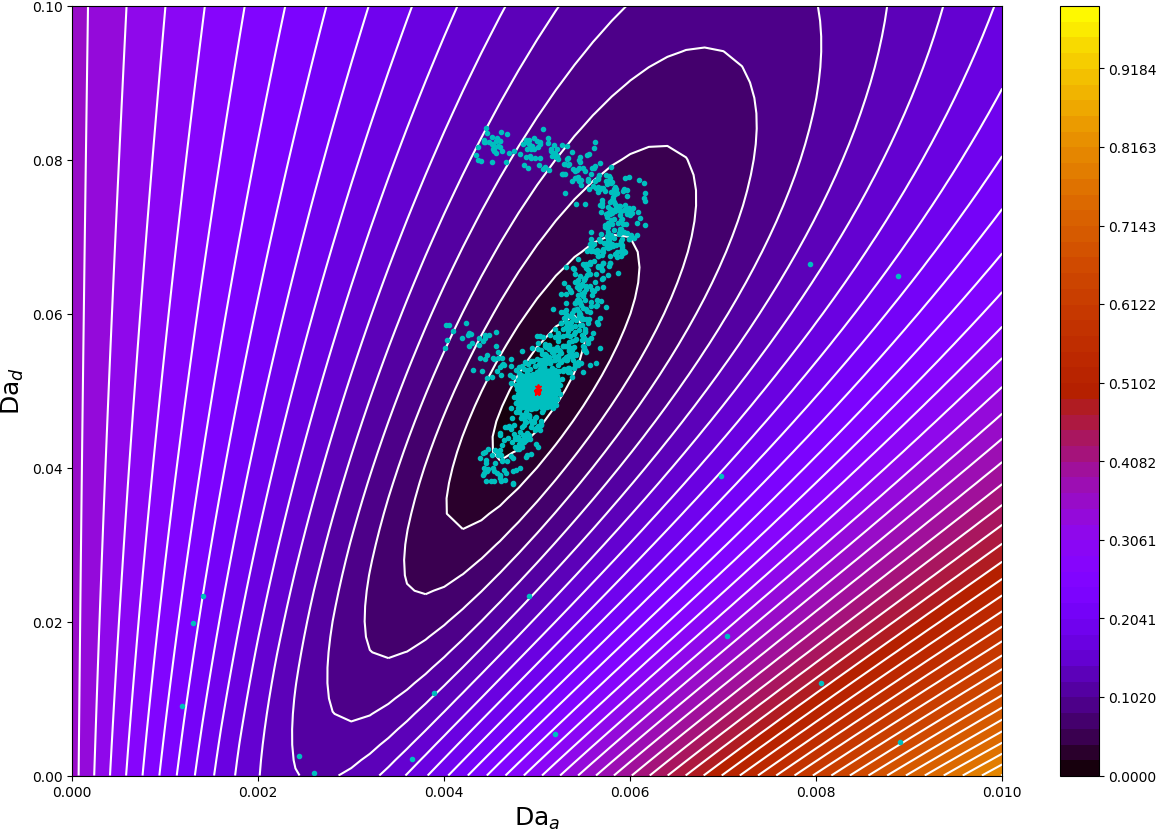}}
\caption{2 best and 3 perspective locations.}
\label{dd_mbc_alot}
\end{minipage}
\hfill
\begin{minipage}[t]{0.45\linewidth}
\center{\includegraphics[width=1.0\linewidth]{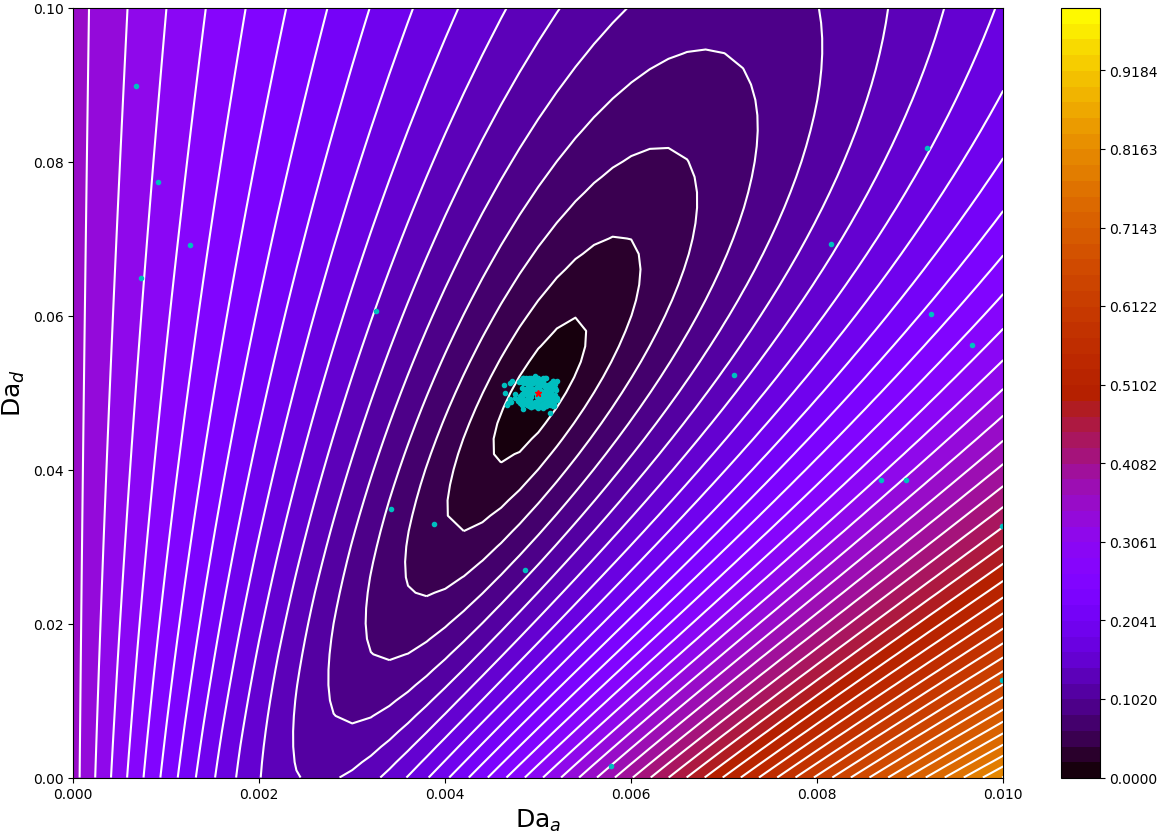}}
\caption{1 best location.}
\label{dd_mbc_one}
\end{minipage}
\end{figure}
\FloatBarrier

For this case, we specially took 2 best locations and 3 perspective ones to show how the bee colonies all converge on one point. As you can see in Fig.\ref{dd_mbc_alot} all the bees merged into one point. The number of direct problem evaluations is also not so big. For this case, the functional we have has a good smooth shape, and we have only one global extremum. If we have such a priori information about functional, then we can run our algorithm with the condition that there is only 1 best location and 0 perspective locations, other parameters are the same. For 1 best location we got following result: \\

J(0.0050001, 0.0499864) = 6.4601e-05 \\
Total direct problem computing (times): 220 \\

The algorithm gives us the exact coordinates of the extremum with substantially small numbers of direct problem evaluations (see Fig.\ref{dd_mbc_one}). At the same time, now it looks more like a multistage parameter identification.

\section{Conclusion}

The Modified Bee Colony Algorithm for identification of unknown adsorption and desorption rates in Langmuir isotherm are presented in conjunction with pore scale simulation of reactive flow. Simulation results show that MBC algorithm can be efficiently employed to solve the multimodal engineering problems with high dimensionality. 

\begin{enumerate}
 \item 
The 2D mathematical model of the direct problem includes steady state Stokes equations, and convection--diffusion equation supplemented with Robin type boundary conditions accounting for adsorption and desorption. Langmuir isotherm describe the kinetics. A simple pore scale geometry described by periodic arrangement of cylindrical obstacles, is considered for illustration of the identification procedure. The key dimensionless parameters are specified. The approach is applicable for wide range of microgeometries and process parameters.

 \item 
The numerical solution is based on triangular grids and FEM with Taylor and Hood elements. Spatial and temporal investigations are performed.

 \item 
Mass transport is simulated for as given velocity field (one way coupling). The numerical solution is based on FEM with piecewise linear elements. Crank-Nikolson scheme is used in the time discretization. Sensitivity studies are carried out to investigate the influence of different parameters on the reactive transport through the porous media.

 \item 
The Modified Bee Colony Algorithm are presented and tested. A comparative work is done with the previous work on the identification of parameters in diffusion dominated case.
\end{enumerate} 

\section*{Acknowledgements}

The work was supported by Mega-grant of the Russian Federation Government (N 14.Y26.31.0013), and by grant from the Russian Foundation for Basic Research (project 19-31-90108).


\end{document}